\documentclass[letterpaper,twocolumn,10pt]{article}
\usepackage{zhanggroup}

\usepackage{tikz}
\usepackage{amsmath}
\usepackage{graphicx}
\usepackage{hyperref}
\usepackage{booktabs}
\usepackage{xspace}
\usepackage{subcaption}
\usepackage{multirow}
\usepackage[capitalize]{cleveref}
\usepackage[absolute]{textpos}

\newcommand{\mypara}[1]{\noindent\textbf{#1.}}

\def\method{{\sc FakePCD}\xspace}
\def\closeworld{close-world attribution\xspace}
\def\openworld{open-world attribution\xspace}
\def\PerShape{single-shape\xspace}
\def\ShapeExtended{multiple-shape\xspace}

\begin{document}

\begin{textblock}{12}(2,1)
\centering
To Appear in the 19th ACM ASIA Conference on Computer and Communications Security, July 1-5, 2024.
\end{textblock}

\date{}

\title{\method: Fake Point Cloud Detection via Source Attribution}

\author{
Yiting Qu\textsuperscript{1}\ \ \
Zhikun Zhang\textsuperscript{1}\ \ \
Yun Shen\textsuperscript{2}\ \ \
Michael Backes\textsuperscript{1}\ \ \
Yang Zhang\textsuperscript{1}
\\
\\
\textsuperscript{1}\textit{CISPA Helmholtz Center for Information Security} \ \ \ 
\textsuperscript{2}\textit{Netsapp}
}

\maketitle

\begin{abstract}
To prevent the mischievous use of synthetic (fake) point clouds produced by generative models, we pioneer the study of detecting point cloud authenticity and attributing them to their sources.
We propose an attribution framework, \method, to attribute (fake) point clouds to their respective generative models (or real-world collections).
The main idea of \method is to train an attribution model that learns the point cloud features from different sources and further differentiates these sources using an attribution signal.
Depending on the characteristics of the training point clouds, namely, sources and shapes, we formulate four attribution scenarios: close-world, open-world, \PerShape, and \ShapeExtended, and evaluate \method's performance in each scenario.
Extensive experimental results demonstrate the effectiveness of \method on source attribution across different scenarios.
Take the \openworld as an example, \method attributes point clouds to known sources with an accuracy of 0.82-0.98 and to unknown sources with an accuracy of 0.73-1.00.
Additionally, we introduce an approach to visualize unique patterns (fingerprints) in point clouds associated with each source.
This explains how \method recognizes point clouds from various sources by focusing on distinct areas within them.
Overall, we hope our study establishes a baseline for the source attribution of (fake) point clouds.\footnote{Our code is available at \url{https://github.com/YitingQu/FakePCD}}
\end{abstract}

\section{Introduction}
\label{section:introduction}

Point clouds are a set of data points in 3D space to represent 3D objects or scenes~\cite{LZKK21,GWHLLB21}.
They are commonly collected using 3D laser scanners~\cite{DLYXZLWDFHS20,HYXRGWTM20} and LiDAR (light detection and ranging) technology~\cite{YLU19,DUKVQMF11}.
Due to their ability in preserving geometric information in 3D space, point clouds have been widely used for 3D modeling scenarios, e.g., autonomous driving~\cite{GWHLLB21,LMZLCCL21,NBR21}, robotics~\cite{DDSNSC17}, and shape synthesis~\cite{SCCL20,GLHZF21}.

\begin{figure}[t]
\centering
\begin{subfigure}{0.4\columnwidth}
\includegraphics[width=\columnwidth]{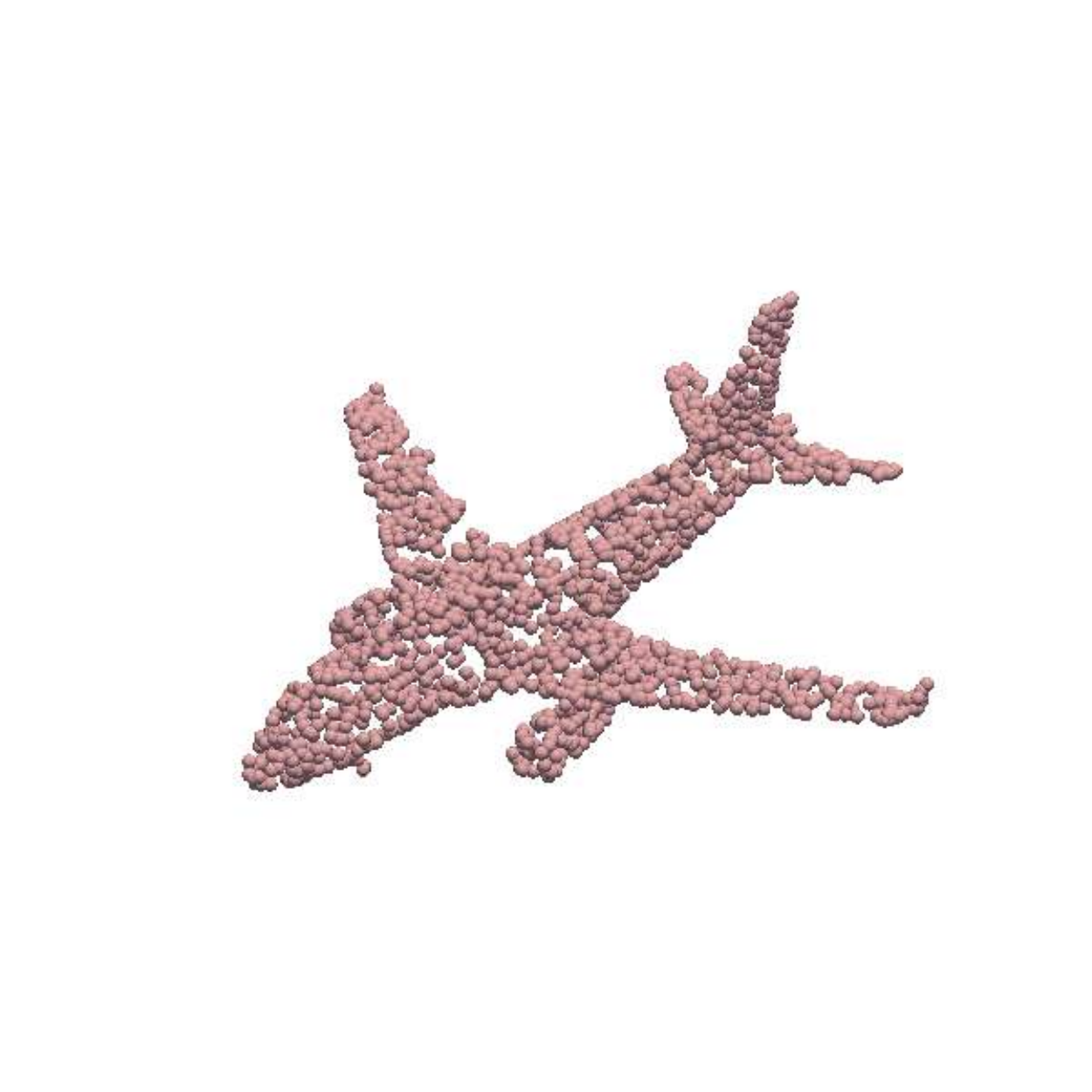}
\caption{Real}
\end{subfigure}
\begin{subfigure}{0.4\columnwidth}
\includegraphics[width=\columnwidth]{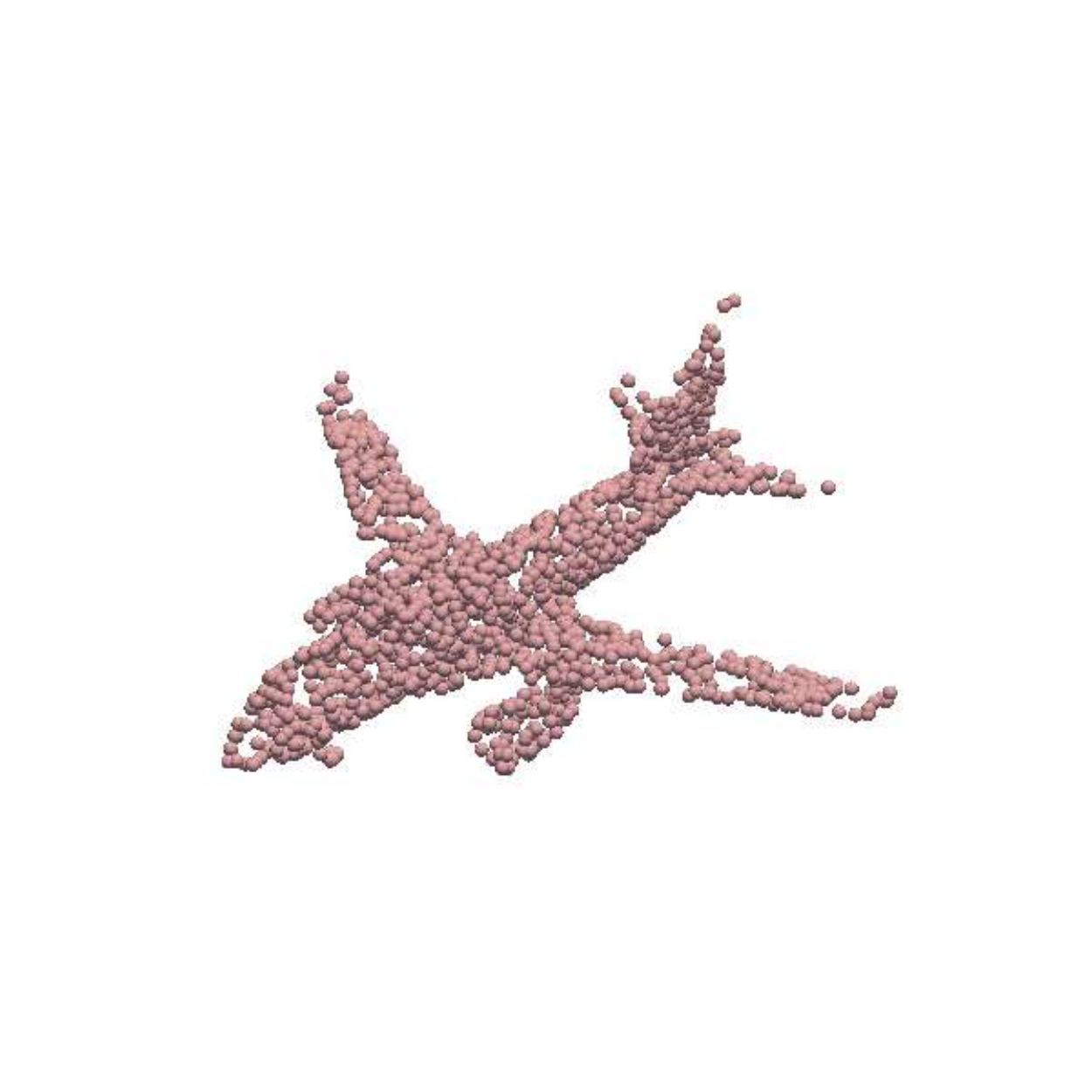}
\caption{Fake}
\end{subfigure}
\caption{Examples of a real point cloud (airplane) and a fake point cloud. 
The fake point cloud is generated with Progressive Deconvolution Generation Network (PDGN)~\cite{HXXQY20}.}
\label{figure:real_fake_examples}
\end{figure}

Recently, deep learning techniques have accelerated the development of point cloud generative models~\cite{HXXQY20,YHHLBH19,KYLH21,CYAHBSH20,LH21}.
These models can generate synthetic point clouds with different \emph{shapes}, with each shape representing an object, such as an airplane, car, or chair.
These synthetic (fake) point clouds are resembling the real-world ones, and it is difficult to distinguish them with human eyes~\cite{YHHLBH19,CYAHBSH20,LH21,KYLH21,HXXQY20} (see \autoref{figure:real_fake_examples}).
This advancement eliminates the need for costly manual point cloud acquisition, thereby improving the training of various models in classification~\cite{QSMG17,WSLSBS19}, segmentation~\cite{DUKVQMF11,DDSNSC17}, and object detection~\cite{YLU19}.

However, a coin has two sides.
Despite their utility, synthetic point clouds might also be used maliciously.
For instance, the data collector expects point clouds scanned by terrestrial laser scanning (TLS), the contractor, instead, provides synthetic point clouds to save cost,  which may be inferior in quality and introduce the potential risk~\cite{WBTTM17,LKNKKLHRVVKIZJH17}.
In addition, an adversary could leverage LiDAR spoofing attacks~\cite{CXCZPRCFM19} to inject synthetic point clouds (e.g., fake aircraft) into the UAV's LiDAR sensors.
This can render the Intelligence Surveillance Reconnaissance (ISR) systems ineffective (i.e., collecting misleading military intelligence) during real-world operations.
The above work exemplifies how synthetic point clouds can be generated and utilized for malicious purposes.

\mypara{Research Problem}
To reduce the potential security risks of synthetic point clouds, it is imperative to detect the authenticity of point clouds, i.e., the point clouds are collected from the real world or generated by different models.
Identifying synthetic data from generative models allows developers to implement protective measures, such as watermarking, to prevent misuse.
To this end, we pioneer study into the \emph{source attribution}~\cite{YDF19} of point clouds: \emph{attributing point clouds to the corresponding sources~\cite{ZZSP21}}, i.e., collected from the real world or generated by a specific model.
For instance, given two point clouds of the \emph{shape} airplane displayed in \autoref{figure:real_fake_examples}, we aim to identify whether the point cloud is real or generated with a particular model, i.e., PDGN~\cite{HXXQY20}.

\mypara{Challenges}
This attribution problem leads to non-trivial challenges.
First, to attribute data to their sources, existing efforts~\cite{YDF19} on synthetic images often assume that all sources at both training time and testing time are known, i.e., \closeworld.
However, in the real-world scenario, point clouds from unknown sources are often fused with those of known sources at the test time.
We need to identify point clouds from known sources and recognize those from unknown sources.
We term this more challenging scenario as \openworld (see formulation in \autoref{section:problem_statement}).
Secondly, different from synthetic images that can be sampled with one model, e.g., StyleGAN~\cite{KLA19}, for synthetic point clouds, training a new generative model is required for every shape.
This suggests that with a single attribution model, we typically trace the source of point clouds if they share the same shape.
However, enumerating all shapes can be resource-consuming.
Similar to unknown sources, point clouds of new (unseen) shapes are often fused with seen shapes during source attribution.
The above two points bring challenges to the source attribution of point clouds and are investigated in our study.

\mypara{Our Work}
We propose a (synthetic) point cloud detection and attribution framework, \method, to verify the authenticity of point clouds and trace their sources.
The idea is intuitive.
We first learn the distribution of point clouds from different sources and extract point cloud features.
Then, given a new point cloud, we calculate the feature distance and assign it to the specific source based on a threshold-based criterion.
Specifically, \method comprises three stages: close-world pre-training, open-world pre-training, and threshold-based assignment.
We train the encoder to extract point cloud features in a fully supervised manner in the first stage to solve the close-world attribution problem.
Based on the pre-trained encoder, we train the network in a supervised contrastive manner in the second stage to learn more generalizable features.
Finally, in the threshold-based assignment stage, we measure the distances of a testing example and each type of \emph{anchors} (fixed point clouds from known sources) as the attribution signal.
We classify a point cloud as an unknown source if the attribution signal is stronger than a threshold.
Otherwise, we attribute it to the closest known source.

To evaluate \method's performance, we construct a dataset of 99K point clouds from six sources (five generative models + real-world collection), covering six common shapes, namely, \emph{Airplane}, \emph{Car}, \emph{Chair}, \emph{Bench}, \emph{Cabinet}, and \emph{Lamp}.
We then categorize these point clouds along two dimensions: known and unknown sources, seen and unseen shapes.
Point clouds of known sources and seen shapes serve as training data for \method to learn point cloud features.
In \closeworld, we apply \method to attribute point clouds to known sources.
In \openworld, we evaluate \method's performance both in identifying point clouds from known sources and recognizing ones from unknown sources.
In both attribution scenarios, we further study the \method's generalizability to attributing point clouds of unseen shapes.

Additionally, we explain how \method attributes point clouds using \emph{critical points}~\cite{QSMG17}, \emph{a set of points within a point cloud that can significantly affect a model's prediction}.
Using these critical points, we introduce a novel approach to visualize the unique patterns in point clouds from different sources.
Specifically, we calculate critical points for each source and project them to the x-y plane as depth images.
By stacking a certain number of depth images from the same source, we observe unique patterns associated with each generative model, namely, fingerprints.

\mypara{Results}
Experimental results show that \method perfectly identifies point clouds from known sources in \closeworld.
It also achieves an accuracy ranging from 0.72 to 0.79 when attributing point clouds of unseen shapes, which are not included in its training data.
For \openworld, \method attributes point clouds to known sources with 0.82-0.98 accuracy and to unknown sources with 0.73-1.00 accuracy.
In this scenario, \method demonstrates generalizability when attributing point clouds of unseen shapes, especially from known sources, with 0.58-0.84 accuracy.
Interestingly, we find that if an unseen shape closely resembles seen shapes in the training data, for example, Bench (unseen) being similar to Chair (seen), then the attribution performance for Bench tends to be higher compared to other unseen shapes.
Furthermore, we visualize the unique patterns associated with each source with examples of Airplane, Car, and Chair.
We find that \method focuses on different areas within point clouds depending on their sources.

\mypara{Contributions}
Our contributions are summarized as follows:

\begin{itemize}
\item We provide the first work on point cloud authenticity and source attribution both in the close-world and open-world settings.

\item We propose \method to attribute a given point cloud to its source and recognize open-world examples. 
We also explore \method's generalizability in attributing point clouds of unseen shapes.

\item We introduce an approach to visualize the unique patterns (fingerprints) within point clouds for different sources. This visualization illustrates how \method attributes point clouds back to their respective sources.
\end{itemize}

\section{Preliminary}
\label{section:preliminary}

\subsection{Point Clouds}
\label{subsection:point_clouds_preliminary}

A point cloud is a set of discrete points $O$.
Each point $o \in O$ is a tuple that represents the $x$, $y$, and $z$ coordinates.
These points are recorded and connected to form the surface of a 3D object in Euclidean space (see \autoref{figure:real_fake_examples} for illustration of point clouds).
They can be captured through various methods, such as laser scanning or photogrammetry (e.g., LiDAR and Kinect). 
Due to their ability to preserve geometric information in 3D space, they are commonly used in fields such as computer graphics~\cite{GWHLLB21}, manufacturing~\cite{SDDRKB17}, robotics~\cite{PCS15}, autonomous driving~\cite{LMZLCCL21}, remote sensing~\cite{LIPMGSW10}, and architecture~\cite{LZKK21}.
The common research tasks on point clouds include classification (categorizing points in a point cloud into semantic classes, e.g., aircraft)~\cite{QSMG17}, segmentation (dividing a point cloud into smaller semantic pieces, e.g., wings)~\cite{QSMG17}, reconstruction (creating a 3D mesh or surface from a point cloud)~\cite{VD01}, registration (aligning multiple point clouds to a single coherent representation)~\cite{PCS15}, etc.

\subsection{Point Cloud Generative Models}
\label{subsection:generative_models_preliminary}

Inspired by the success of generative models in the image domain~\cite{SK19}, generative models for point clouds have emerged as a type of machine learning algorithms that can generate synthetic 3D point clouds~\cite{HXXQY20,YHHLBH19,KYLH21,CYAHBSH20,LH21}.
These models are trained on large datasets of real-world point clouds to learn the underlying patterns and distributions that govern the generation of 3D shapes.
The main challenge in this field is to synthesize high-quality point clouds that are diverse and faithful to the training data.
To address the challenge, several types of generative models for point clouds have been developed, including Variational Autoencoders (VAEs)~\cite{KYLH21}, Generative Adversarial Networks (GANs)~\cite{HXXQY20}, normalizing flow~\cite{YHHLBH19} and diffusion models~\cite{LH21}.
These models tend to focus on improving the quality of generated shapes, enhancing efficiency, and extending the models to handle more complex data distributions.
We refer the audience to Shi et al.~\cite{SPXLS22} for a comprehensive survey of this research area.

\section{Threat Model}

\subsection{Adversary's Goal}
\label{subsection:adversary_goal}

In this paper, we consider adversaries that have the expertise to train and run point cloud generative models.
Their goal is to produce fake point clouds with generative models and use them for malicious purposes.
The application scenarios of these fake point clouds are introduced in \autoref{subsection:application_scenarios}).
For instance, if their goal is to replace the original 3D design with generated point clouds to disrupt the manufacturing process, the adversary can launch (or outsource) cyber attacks to breach the targeted manufacturing system.

\subsection{Application Scenarios}
\label{subsection:application_scenarios}

We consider three real-world application scenarios where point clouds might be used maliciously and source attribution can be an effective approach to identify the misuse.
Their respective details are outlined below.

\begin{itemize}
\item \textbf{Inauthentic Point Cloud Scanning}.
In this scenario, the data collector anticipates genuine point clouds acquired through terrestrial laser scanning (TLS).
The contractor, instead of scanning the real objects, provides artificially generated point clouds to save costs.
These fake point clouds may be inferior in quality and can lead to models learning imprecise distributions~\cite{WBTTM17,LKNKKLHRVVKIZJH17}.

\item \textbf{Manufacturing Disruption}.
In this scenario, the adversary uses the generated point clouds to disrupt the manufacturing process. 
Take additive manufacturing (AM), commonly known as 3D printing in the consumer space, for instance. 
It relies on accurate 3D models for the printing process, allowing the machine to build the object layer by layer~\cite{GRSKRSK21}. 
Due to the digital nature of AM (a network that connects computers and robots), the adversary may attack the AM supply chain via cyber-attack and replace the original design with the generated point clouds.
Note that these generated point clouds are visually close to the real designs, which may not be noticed by operators due to automation.
Nonetheless, AM processes are sensitive to small deviations from the design~\cite{RLRKW15}.
Consequently, the adversary can compromise the quality of the print parts and introduce undesired deficiencies. 
This practical scenario has been discussed in Ye et al.~\cite{YLTK20}

\item \textbf{Counter Intelligence Surveillance and Reconnaissance}. 
In this scenario, the adversary leverages spoofing attacks on LiDAR sensors~\cite{CXCZPRCFM19,SCCM20} to inject generated point clouds (e.g., fake aircraft) in the LiDAR field of view of unmanned aerial vehicles (UAVs). 
This attack undermines Intelligence Surveillance Reconnaissance (ISR) systems, misleading real-world operations with falsified military intelligence. 
Here, we do not consider the case that the adversary may remove point regions from LiDAR readings~\cite{CBNSMR22} and, consequently, UAV may not identify the surveillance targets.
\end{itemize}

\subsection{Attribution Capability}
\label{subsection:attribution_capability}

To reliably trace the generative models that originally generated the suspicious point clouds, we assume the following attribution capabilities.

\begin{itemize}
\item \textbf{Access to Authentic Point Clouds}.
The attribution framework has access to adequate point clouds that are obtained from real-world objects.
Note that \emph{shape} information (e.g., Car, Airplane) naturally comes with the point clouds.
We denote these point clouds as \emph{real} point clouds for ease of presentation in the rest of the paper.

\item \textbf{Access to Existing Generative Models}.
The attribution framework has access to a list of known point cloud generative models.
These models are our attribution targets.
We denote these models as \emph{sources} for ease of presentation in the rest of the paper.
Note that the attribution framework can trivially generate point clouds from these sources to form its training data.
\end{itemize}

In \autoref{subsection:exp_open_world}, we show that these two capabilities are sufficient for the attribution framework to attribute point clouds to even unknown sources.
This makes our attribution framework more practical in the real world.

\section{Attribution Problem Formulation}
\label{section:problem_statement}

To detect whether a point cloud is generated or collected from the real world, we solve this problem by attributing point clouds to the corresponding sources.
As mentioned in \autoref{section:introduction}, two major factors affect the attribution difficulty - settings (i.e., \closeworld or \openworld) and shapes.
We outline them below.

\mypara{Settings}
Formally, assume there are $\mathcal{K}$ known sources labeled as $\{G_1, G_2,..., G_\mathcal{K}\}$ and one additional Real class labeled as $R$.
Each source has $N$ point clouds sampled from a given shape (e.g., Car).
Given a point cloud $x_i \in X$ and its corresponding source $y_i$, we consider the following settings when conducting the attribution:

\begin{itemize}
\item \textbf{Close-World Attribution:} If $x_i$ is a point cloud with $y_i$ from the existing labels $\{G_1, G_2,..., G_\mathcal{K}, R\}$, the attribution is considered successful if the predicted label $\hat y_i = y_i$.

\item \textbf{Open-World Attribution:} If $x_i$ is a point cloud with $y_i \notin \{G_1, G_2,..., G_\mathcal{K}, R\}$, the attribution is considered successful if $\hat y_i = G_{K+1}$, where $G_{K+1}$ denotes the unknown source.
\end{itemize}

\begin{figure*}
\centering
\includegraphics[width=0.7\textwidth]{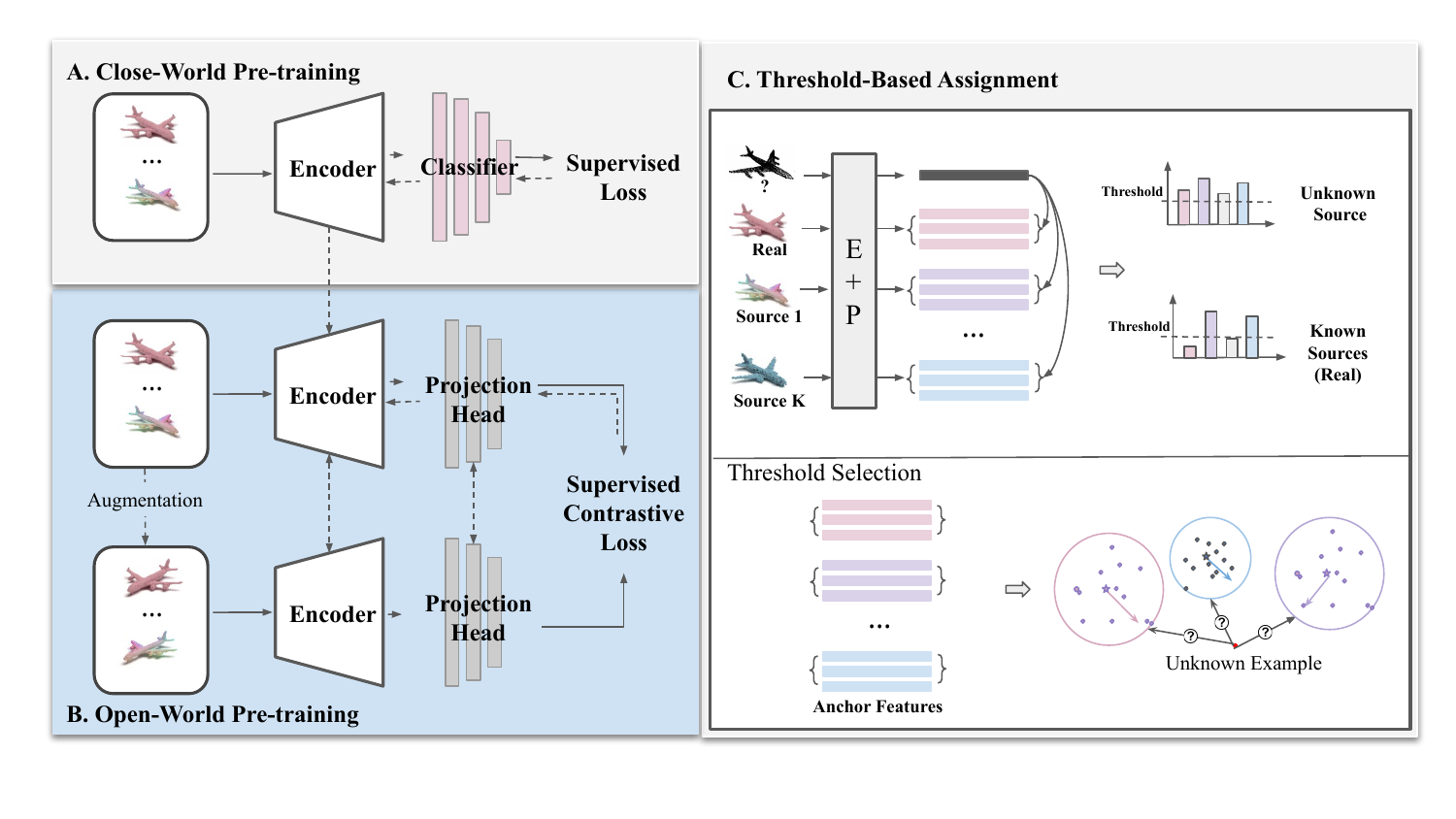}
\caption{Workflow of \method. 
``E+P'' denotes the attribution model comprised of an encoder and a projection head.
In the close-world pre-training stage, we train the attribution model in a fully supervised manner to solve the close-world attribution.
In the open-world pre-training stage, models learn more differentiated point cloud features by supervised contrastive training.
Threshold-based assignment enables us to attribute a point cloud to its source.
By selecting point clouds of one-shape or multi-shape as the training dataset, this framework adapts to all the scenarios introduced in \autoref{section:problem_statement}.}
\label{figure:overview}
\end{figure*}

\mypara{Shape}
The shapes of point clouds $S=\{s_1, s_2, ..., s_m\}$ in the training data also exert a great influence on the attribution outcome.
Intuitively, the attribution performance is likely to be better on the point clouds of seen shapes (in the training data) compared to unseen shapes.
To explore the model's generalizability in attributing point clouds of unseen shapes, \emph{multi-shape} modeling should also be considered.
For each type of setting, we further consider the following two scenarios.

\begin{itemize}
\item \textbf{Single-Shape Scenario:}
The attribution model is trained and tested on $X^{s_i}$, where $X^{s_i}$ denotes the point clouds dataset of a single shape $s_i$.
The attribution model is regarded as \emph{single-shape model}, as it simply attributes point clouds of the same shape, e.g., Airplane.
 
\item \textbf{Multiple-Shape Scenario:}
The attribution model is trained on $S'=\{X^{s_1}, ..., X^{s_l}\}$ and tested on $S''=S' \cup \{X^{s_{l+1}}, ..., X^{s_z}\}$ where $l < z \leq m$. We regard it as \emph{multiple-shape model}, which attributes point clouds of multiple shapes simultaneously including unseen ones.
For instance, our attribution model is trained on Airplane, Car, and Chair.
At test time, we attribute an unseen Bench point cloud to its source.
\end{itemize}

\begin{table}[!t]
\centering
\caption{Summary of attribution scenarios.}
\label{table:attribution_scenario}
\scalebox{0.80}{
\begin{tabular}{p{0.15\columnwidth}|p{0.45\columnwidth}|p{0.45\columnwidth}}
\toprule
    &  Close-World Attribution  & Open-World Attribution \\
\midrule
Single-Shape    & A Single Shape;\newline Known Sources   &  A Single Shape;\newline Known/Unknown Sources  \\
\hline
Multi-Shape     &  Seen/Unseen Shapes;\newline Known Sources  & Seen/Unseen Shapes;\newline Known/Unknown Sources \\
\bottomrule
\end{tabular}
}
\end{table}

\mypara{Summary}
We consider four attribution scenarios in the paper, which are summarized in \autoref{table:attribution_scenario}.
Specifically, we perform close-world attribution and open-world attribution in both the single-shape scenario and multiple-shape scenario respectively.

\section{Attribution Framework}
\label{section:methodologies}

\subsection{Overview}

We present the attribution framework of (synthetic) point clouds, \method.
\autoref{figure:overview} illustrates the workflow of \method.
The main idea of our framework is to train an encoder to capture the feature distance among different sources.
The point cloud is then assigned to a source based on a threshold criterion.
Concretely, the framework is composed of three stages and takes all the aforementioned four scenarios into account. 

\mypara{Close-World Pre-training (\autoref{subsection: encoder_pretraining})}
Our attribution model consists of an encoder $f$ and a classifier $g$, which is trained in a fully supervised manner.
This stage alone can deal with the close-world attribution problem in both \PerShape and \ShapeExtended scenarios in our evaluation.
It also serves as the stepping-stone for \openworld.
The process is illustrated in Part A of \autoref{figure:overview}.

\mypara{Open-World Pre-training (\autoref{subsection: encoder_finetuning})}
In this stage, we employ the supervised contrastive learning approach~\cite{KTWSTIMLK20} to address the point cloud attribution problem.
The attribution model in this stage consists of the same encoder $f$ and a projection head $h$. 
Each point cloud is augmented via point cloud transformations to create its augmented version. 
We then feed both point clouds to the two identical models with sharing weights.
Based on the pre-trained encoder from the first stage, we continue pre-training the encoder $f$ with supervised contrastive loss.
The process is illustrated in Part B of \autoref{figure:overview}.

\mypara{Threshold-Based Assignment (\autoref{subsection: threshold-based assignment})}
We first preserve the open-world pre-trained encoder $f$ and projection head $h$ in the attribution model.
We then build an anchor set that contains a small set of point clouds from known sources.
Given a testing sample, we obtain its feature vector from the attribution model and then measure the distance to each example in the anchor set.
We then average these distances to obtain a mean distance value for each source.
Using a threshold-based criterion (see \autoref{subsection: threshold-based assignment}), we assign the example to one of the existing sources or a new one (i.e., the unknown source).
The process is illustrated in Part C of \autoref{figure:overview}.

\mypara{Note}
\method is a framework.
The users can use or fine-tune different encoders to instantiate their attribution models.
For ease of presentation, we use an attribution model to describe technical details in different stages in this section, then use \method to describe evaluation results in \autoref{section:evaluation}.

\subsection{Close-World Pre-training}
\label{subsection: encoder_pretraining}

We train the attribution model that is composed of an encoder $f$ and a classifier $g$ in a fully-supervised manner.
The parameters of the attribution model (encoder + classifier) are optimized based on the cross-entropy loss.
Employing supervised learning with cross-entropy loss is a primary solution for addressing closed-world classification problems.
We resolve the close-world attribution problem in both \PerShape and \ShapeExtended scenarios in this stage.

\subsection{Open-World Pre-training}
\label{subsection: encoder_finetuning}

Supervised learning generally does not yield good results in open-world attribution, especially under a threshold-based criterion, as it is not designed to handle open-world data. Our initial investigations show that the attribution model, when trained in a fully supervised manner, tends to misclassify point clouds from unknown sources to known ones with overwhelmingly high probabilities (see \autoref{subsection:exp_open_world}).
Supervised contrastive learning~\cite{KTWSTIMLK20}, on the other hand, adapts the self-supervised contrastive approach to the supervised setting.
With this learning paradigm, we first create an augmented version for each point cloud in the training set.
We then feed both original and augmented point clouds to the encoder and projection head to obtain feature vectors.
We optimize the encoder and projection head to maximize the similarity between feature vectors of positive pairs (same labels) while minimizing the similarity for negative pairs (different labels).
We use the supervised contrastive loss (L)~\cite{KTWSTIMLK20} to formulate the optimization process.

\begin{equation}
\label{equation: supervised_contrastive training}
L = \min_{\theta_f, \theta_h}{
\Bigl(
\sum_{x_i \in \mathbf{X}}{\frac{-1}{|\mathbf P(x_i)|}}
\sum_{p \in \mathbf{P}(x_i)}{\log \frac{\exp{(z_i \cdot z_p / \tau)}}{
\sum_{a \in \mathbf A(x_i)}{\exp{(z_i \cdot z_a / \tau)}}}} 
\Bigr)
}
\end{equation}

$\mathbf{X}$ represents the training set that contains original point clouds and their augmented versions.
$\mathbf A(x_i) \equiv \mathbf X \backslash {x_i}$, denotes the set of all examples excluding $x_i$.
$\mathbf P{(x_i)} \equiv \{p \in \mathbf A(x_i): \hat y_p = \hat y_i \}$ is the positive set containing examples with the same label as $x_i$.
$z_i = h(f(x_i))$ is the output of the projection head, i.e., the feature vector of $x_i$.
$\tau$ is the temperature scaling parameter.

We apply three types of augmentations commonly used with point clouds: \emph{translation}, \emph{jittering}, and \emph{rotation}.
Translation moves the origin of coordinates and builds a new coordinate frame.
Jittering adds small Gaussian noise to the coordinates of points.
Rotation turns point clouds with random angles along x, y, and/or z axis.
Practically, when calculating feature vectors, we apply two identical encoders and projection heads, each processing either the original point clouds or the augmented ones.
During the optimization, we keep one pair of the encoder and projection head frozen and only update the other pair (see \autoref{figure:overview}).
Note that the encoder can be randomly initialized or initialized with the checkpoint we obtained from the close-world pre-training stage.
Finally, the encoders learn differentiated and generalized point cloud features.
Based on these feature vectors, we further introduce a threshold-based criterion to assign point clouds to their respective sources.

\subsection{Threshold-Based Assignment}
\label{subsection: threshold-based assignment}

We design the following threshold-based criterion to assign a testing example to either a specific known source or an unknown source. 
That is, \emph{if its distances to all known sources are larger than a threshold, we recognize the example as a sample from an unknown source; otherwise, we assign it to the closest known source.}
To this end, we first randomly sample point clouds from $\mathcal{K}$ known sources, denoted as \emph{Anchor Set}, with each source containing $N$ point clouds.
These anchor point clouds are then forwarded to the attribution model and obtain the feature vectors.
Given a testing point cloud, $x_i$, we measure the Euclidean distance between its feature vector ($z_i$) and all feature vectors derived from the anchor set.
We then group these distance values into $\mathcal{K}$ clusters ($\mathcal{C}_j$) based on their sources and calculate a mean distance value, $d_j(z_i)$, for each source ($y_j$).
We compare these mean values to a selected threshold to attribute the point clouds.

\begin{equation}
\label{equation: distance_calculation}
d_j(z_i) = \frac{1}{|\mathcal{C}_j|}\sum_{x_j \in \mathcal{C}_j}\|z_i - z_j\|_2
\end{equation}

\mypara{Threshold Selection}
We select thresholds based on the intra-cluster distances of $\mathcal{K}$ clusters in the anchor set.
Intra-cluster distances are the distances between examples within the cluster to the cluster centroid.
For example, for point clouds in the cluster ($\mathcal{C}_j$), we calculate the Euclidean distance between the centroid and all point clouds in this cluster.
We calculate the intra-cluster distances for $\mathcal{K}$ clusters and obtain $\mathcal{K}$ distance sequences, each representing the compactness of a specific source.
An ideal threshold should be identified within these sequences.
It should enable the assignment of a testing example from known sources to one of the $\mathcal{K}$ clusters, while also assigning a testing example from unknown sources to a new source.
However, the threshold selection is non-trivial as these $\mathcal{K}$ distance sequences might have different distributions.
To address this, we introduce another parameter, $P$-percentile, to control the distance threshold.
The distance threshold is the smallest $P$-percentile value among all distance sequences.
As such, threshold selection is transformed into a percentile selection problem.
Note, multiple thresholds can be selected, with one corresponding to each cluster.
Here, we use a unified threshold for simplicity, and this approach has already proven effective in \autoref{section:evaluation}.

\section{Evaluation}
\label{section:evaluation}

\subsection{Experimental Setup}
\label{subsection:experimental_setup}

\mypara{Real Point Cloud Dataset}
For real-world collected point clouds, we use the dataset ShapeNetCore, a subset of ShapeNet~\cite{CFGHHLSSSSXYY15}.
It contains 55 object shapes such as Airplane, Car, Chair, Bench, Cabinet, and Lamp.
We specifically choose these six shapes because they provide a sufficient number of samples in the dataset, ranging from 1,543 to 7,497 per shape.
In contrast, other shapes are underrepresented with only hundreds of samples, making it hard to construct a sizable dataset.
Each point cloud in the dataset is described by 15,000 points.
To reduce computation complexity, we follow the previous studies~\cite{YHHLBH19,KYLH21,QSMG17,WSLSBS19} and downsample each point cloud to 2,048 points.

\mypara{Syntheic Point Cloud Dataset}
We generate synthetic point clouds using five generative models: PointFlow~\cite{YHHLBH19}, ShapeGF~\cite{CYAHBSH20}, Diffusion~\cite{LH21}, PDGN~\cite{HXXQY20}, and SetVae~\cite{KYLH21}.
These models cover a wide spectrum of generative techniques including Variational Autoencoders (VAEs)~\cite{KYLH21,CYAHBSH20}, Generative Adversarial Networks (GANs)~\cite{HXXQY20}, normalizing flow~\cite{YHHLBH19} and diffusion~\cite{LH21}.
We train each generative model using the real point clouds and generate 4,000 point clouds for Airplane, Car, and Chair, and 1,000 point clouds for Bench, Cabinet, and Lamp.

\begin{itemize}
\item {\textbf{PointFlow}}.
We use PointFlow's official implementation.\footnote{\url{https://github.com/stevenygd/PointFlow}}
We set 128 as the feature dimension in the encoder and 512 as the hidden dimension in the CNF decoder.
We train the generative models for 4,000 epochs with the Adam optimizer and set the initial learning rate to 0.002 which decays linearly after 2,000 epochs.
When sampling generative examples for 6 shapes, we generate shapes and points conditioned on the shape from Gaussian priors.

\item {\textbf{ShapeGF}}.
We use ShapeGF's official implementation.\footnote{\url{https://github.com/RuojinCai/ShapeGF}}
In detail, we first train the auto-encoder models for distribution learning for 2,000 epochs and then train the corresponding generators for 5,000 epochs.
Following the default configuration, the auto-encoder adopts 128 as the dimension of the latent code, and the generator takes a 256-dimensional vector from the Gaussian distribution and outputs the final 256-dimensional latent code.
Adam optimizer is used with a learning rate of 0.0001.
We acquire point clouds of six shapes individually by sampling the corresponding generators.

\item {\textbf{Diffusion}}.
We generate point clouds based on the officially released code.\footnote{\url{https://github.com/luost26/diffusion-point-cloud}}
We select the flow-based model and set the latent dimension as 256.
We then use the Adam optimizer to train the diffusion model for 80,000 iterations with an initial learning rate of 0.002.
When generating new point clouds, we sample from the generative model with a prior drawn from a Normal distribution.

\item {\textbf{PDGN}}.
We use PDGN's official implementation.\footnote{\url{https://github.com/fpthink/PDGN}}
PDGN adopts a progressive GAN model with multiple generators and discriminators.
We use multiple resolutions of (256, 512, 1024, and 2048) point clouds from a 128-dimensional latent vector for the generator based on the original setting.
Four PointNet-like modules serve as the discriminators.
We train the model for 600 epochs with Adam optimizer and an initial learning rate of 0.0001.
In the generating process, we sample point clouds by feeding noise vectors from the Gaussian distribution to the generator and record the results at the highest resolution.

\item {\textbf{SetVae}}.
We use the official implementation of SetVae.\footnote{\url{https://github.com/jw9730/setvae}}
SetVae model has an encoder with \emph{Induced Set Attention Blocks} and a generator with \emph{Attentive Bottleneck Layers}.
The parameters in these components are shared during both the generation and inference processes.
We train the model for 8,000 epochs with the Adam optimizer and an initial learning rate of $10^{-3}$ that linearly decays to 0 after halfway through the training period.
\end{itemize}

\mypara{Dataset Overview}
Overall, we construct a dataset consisting of 99,280 point clouds from six sources (real-world collection + five generative models), covering six common shapes: Airplane, Car, Chair, Bench, Cabinet, and Lamp.
We use the 3D visualization toolkit PyVista\footnote{\url{https://docs.pyvista.org/}} for point cloud rendering.
\autoref{figure:generative_examples} in the Appendix displays generated point cloud examples of six shapes from each source.

\mypara{Experimental Settings}
We split the above dataset into training and test sets at a ratio of 0.6 for each source and shape.
Furthermore, we leave out PDGN and SetVae as unknown sources and take the remaining four as known sources.
Regarding the encoder, we adopt the feature extractor in PointNet~\cite{QSMG17} and DGCNN~\cite{WSLSBS19}.
The classifier and projection head connected to the encoder is a multilayer perceptron (MLP).
For the \PerShape scenario, we train the attribution model on the Airplane, Car, and Chair independently and only test on the point clouds of the same shape.
For the \ShapeExtended scenario, we train on three mixed shapes: Airplane, Car, and Chair, and test on point clouds from all six shapes.
Other experimental settings are outlined in their respective evaluations.

\mypara{Evaluation Metrics}
We measure the attribution performance for tracing both known and unknown sources, using the metrics of \emph{accuracy} and \emph{F1 score}.

\subsection{Close-World Attribution}
\label{subsection:exp_close_world}

\mypara{Setup}
The attribution model in this scenario is composed of an encoder and a classifier.
We employ a 3-layer MLP with (512, 256, $\mathbf{K}$) neurons respectively as the classifier, where $\mathbf{K}=1+3$ with the Real source and three types of generative models (PointFlow, Diffusion, and ShapeGF).
The above selection of known sources is consistent with the \openworld.
The model is trained on 200 epochs at a batch size of 32.
SGD is adopted with a learning rate of 0.1 and momentum of 0.9 to optimize the training procedure.

\mypara{Single-Shape Results}
We calculate the accuracy and F1 score when \method attributes point clouds of a single shape, i.e., Airplane, Car, or Chair.
We find that both \method-PointNet and \method-DGCNN perfectly attribute point clouds of all three shapes (accuracy=1, F1 score=1).
This implies that the feature differences of point clouds from known sources can be well captured.

\begin{figure}[!t]
\centering
\begin{subfigure}{0.9\columnwidth}
\includegraphics[width=\columnwidth]{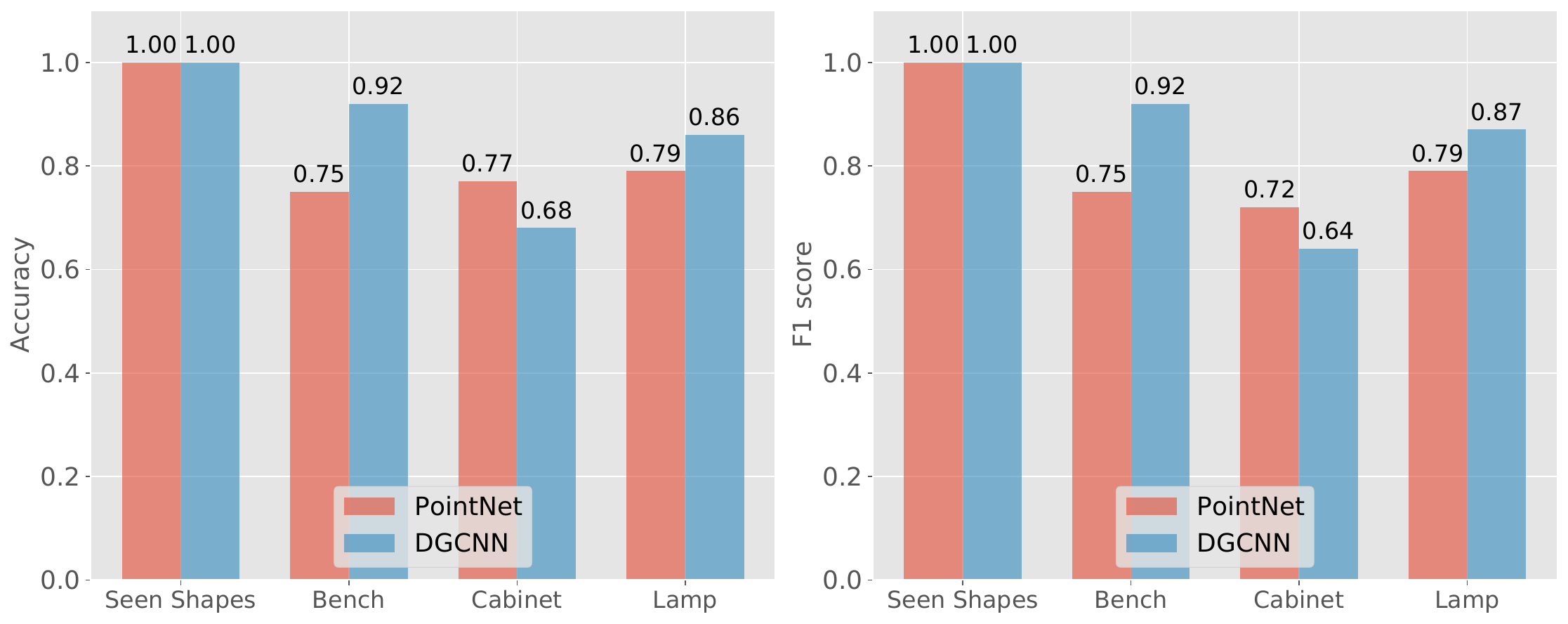}
\end{subfigure}
\caption{Attribution performance in \ShapeExtended scenarios when all the sources are included in the model's training set.
We report the accuracy (left figure) and F1 score (right figure).}
\label{figure:close_world}
\end{figure}

\mypara{Multiple-Shape Results}
We further test the performance of the above models on point clouds of unseen shapes, i.e., Bench, Cabinet, or Lamp.
\autoref{figure:close_world} shows that \method with both encoders still perfectly attribute point clouds of three seen shapes.
\method also achieves an accuracy between 0.68 and 0.92, and an F1 score ranging from 0.64 to 0.92 on unseen shapes.
The results imply that the learned distinction among different sources can be transferred to unseen shapes.
This also suggests that building a unified attribution model that is effective across various shapes is promising, especially when the training dataset contains a sufficient number of point clouds from diverse shapes.
We also find that the DGCNN backbone outperforms the PointNet backbone in two out-of-shape shapes.
We hypothesize that it is due to the robustness of the Graph Neural Network (GNN) employed by DGCNN, which better preserves the local spatial relationship among points.

\mypara{Takeaways}
\method can reliably attribute seen point clouds to known sources.
Moreover, the underlying model used by \method learns sufficient knowledge during the close-world pre-training stage (see Part A in \autoref{figure:overview}) to attribute unseen shapes to known sources with decent accuracy.

\begin{table*}[t]
\centering
\caption{Attribution performance (accuracy and F1 score) of \method and baselines across three shapes.
``P'' and ``D'' denote the encoder (Pointnet and DGCNN).
``Known'' and ``Unknown'' indicate the attribution performance for known and unknown sources.}
\label{table:open_world}
\setlength{\tabcolsep}{0.5em}
\scalebox{0.8}{
\begin{tabular}{lcccccccc}
\toprule
&\multicolumn{2}{c}{Airplane}&&\multicolumn{2}{c}{Car}&&\multicolumn{2}{c}{Chair}\\
\cline{2-3}\cline{4-5}\cline{6-7}
\hline
\multirow{2}{*}{Method} & Acc & F1 score&& Acc & F1 score&& Acc &F1 score\\
&Known\ Unknown & Known\ Unknown && Known\ Unknown & Known\ Unknown && Known\ Unknown & Known\ Unknown\\
\midrule
Fully supervised&\textbf{0.93}$|$0.00 & \textbf{0.95}$|$0.00 && 0.93$|$0.00 &0.96$|$0.00 && 0.87$|$0.00 & 0.90$|$0.00\\
Projection-based&0.66$|$0.04 & 0.69$|$0.07 && 0.88$|$0.12 &0.90$|$0.20 && 0.88$|$0.73 & 0.93$|$0.84\\
Siamese-P&0.14$|$0.25 & 0.19$|$0.40 && 0.83$|$0.83 &0.90$|$0.90 && 0.88$|$0.80 & 0.92$|$0.90\\
Siamese-D&0.82$|$0.29 & 0.89$|$0.45 && 0.75$|$0.53 &0.75$|$0.69 && \textbf{1.00}$|$0.43 & \textbf{1.00}$|$0.60\\
\method-P&0.82$|$\textbf{0.78} & 0.89$|$\textbf{0.87} && 0.95$|$\textbf{1.00} &\textbf{0.98}$|$\textbf{1.00} && 0.90$|$0.99 & 0.95$|$\textbf{1.00}\\
\method-D&0.82$|$0.73 & 0.90$|$0.85 && \textbf{0.96}$|$0.99 &\textbf{0.98}$|$\textbf{1.00} && 0.96$|$\textbf{1.00} & 0.98$|$\textbf{1.00}\\
\bottomrule
\end{tabular}
}
\end{table*}

\subsection{Open-World Attribution}
\label{subsection:exp_open_world}

\begin{figure}
\centering
\includegraphics[width=0.8\columnwidth]{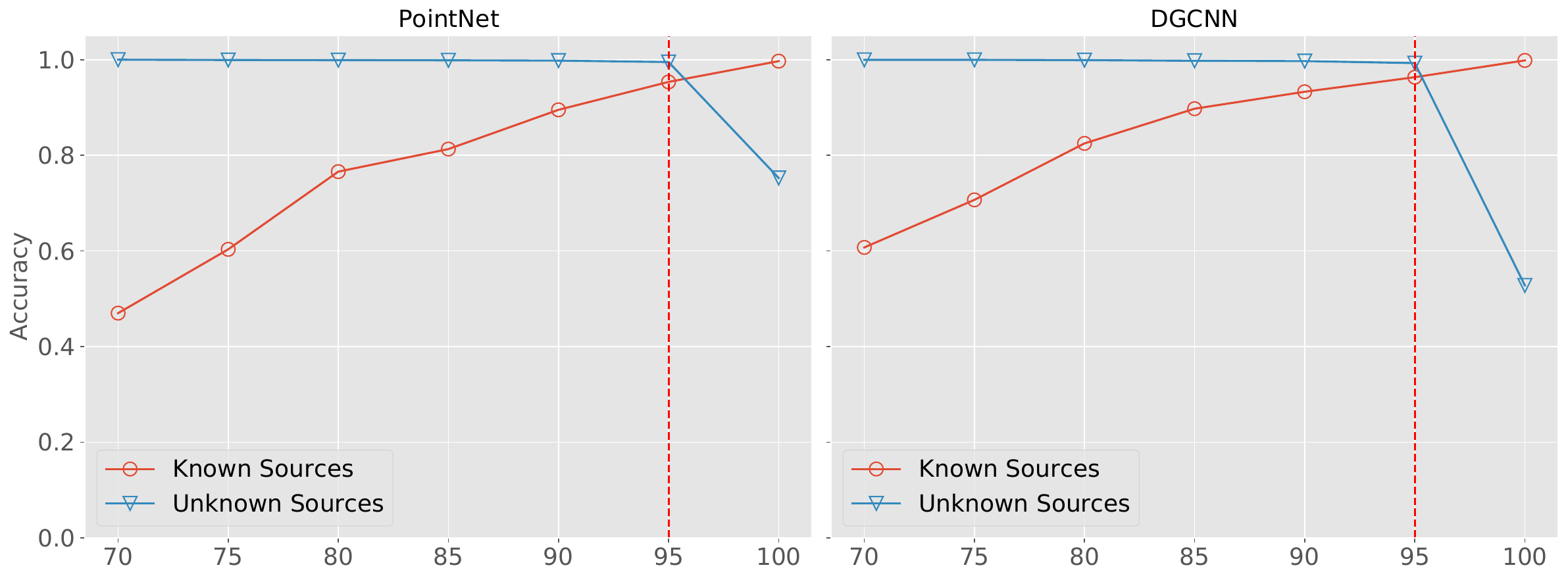}
\caption{$P-$percentile selection when attributing point clouds (Car) using PointNet and DGCNN. 
A larger $P-$percentile leads to increased accuracy for known sources while decreased accuracy for unknown sources. The optimal trade-off is achieved at 95\%.}
\label{figure:select_threshold_example}
\end{figure}

\mypara{Setup}
The attribution model in this scenario consists of an encoder and a projection head.
For the encoder, we initialize the parameters with the checkpoint trained in \closeworld.
For the projection head, we adopt a 2-layer MLP with (512, 128) neurons.
The model is trained for 300 epochs using an early stopping strategy and a batch size of 20.
SGD is employed with a learning rate of 0.1 and a momentum of 0.9 to optimize the training procedure.
In the assignment phase, we first construct the Anchors Set containing 100 random examples from the training set for each known source.
For each unknown source, we use the test set of each shape introduced in \autoref{subsection:experimental_setup}.
To determine the optimal threshold, we randomly sample 100 point clouds from the test set as a validation set and evaluate \method's performances on it across various $P-$percentiles.
Take the shape Car as an example (see \autoref{figure:select_threshold_example}), with a larger $P-$percentile, the accuracy for known sources increases, while it decreases for unknown sources.
The optimal trade-off is achieved at 95\%.
Therefore, we adopt 95\% as the optimal threshold for point clouds of Car and apply it to the remaining examples in the test set.
We report the optimal thresholds for all six shapes in \autoref{table:optimal_thresholds}.
We find that these optimal thresholds range from 70\% to 95\%, tending to be higher for seen shapes (75\%-95\%) and lower for unseen shapes (70\%-75\%).
In practical scenarios, it is advisable to use thresholds within this range when a validation set of unknown sources is not available.

\begin{table}
\centering
\caption{The Optimal $P-$percentile calculated with the validation set of each shape.}
\label{table:optimal_thresholds}
\scalebox{0.8}{
\begin{tabular}{ccccccc}
\toprule
 & Airplane & Car & Chair & Bench & Cabinet & Lamp\\
 \midrule
\method-P & 75\% & 95\% & 90\% & 70\% & 75\% & 70\% \\
\method-D & 85\% & 95\% & 95\% & 70\% & 70\% & 70\% \\
\bottomrule
\end{tabular}
}
\end{table}

\begin{figure*}[t]
\centering
\begin{subfigure}{0.85\columnwidth}
\includegraphics[width=\columnwidth]{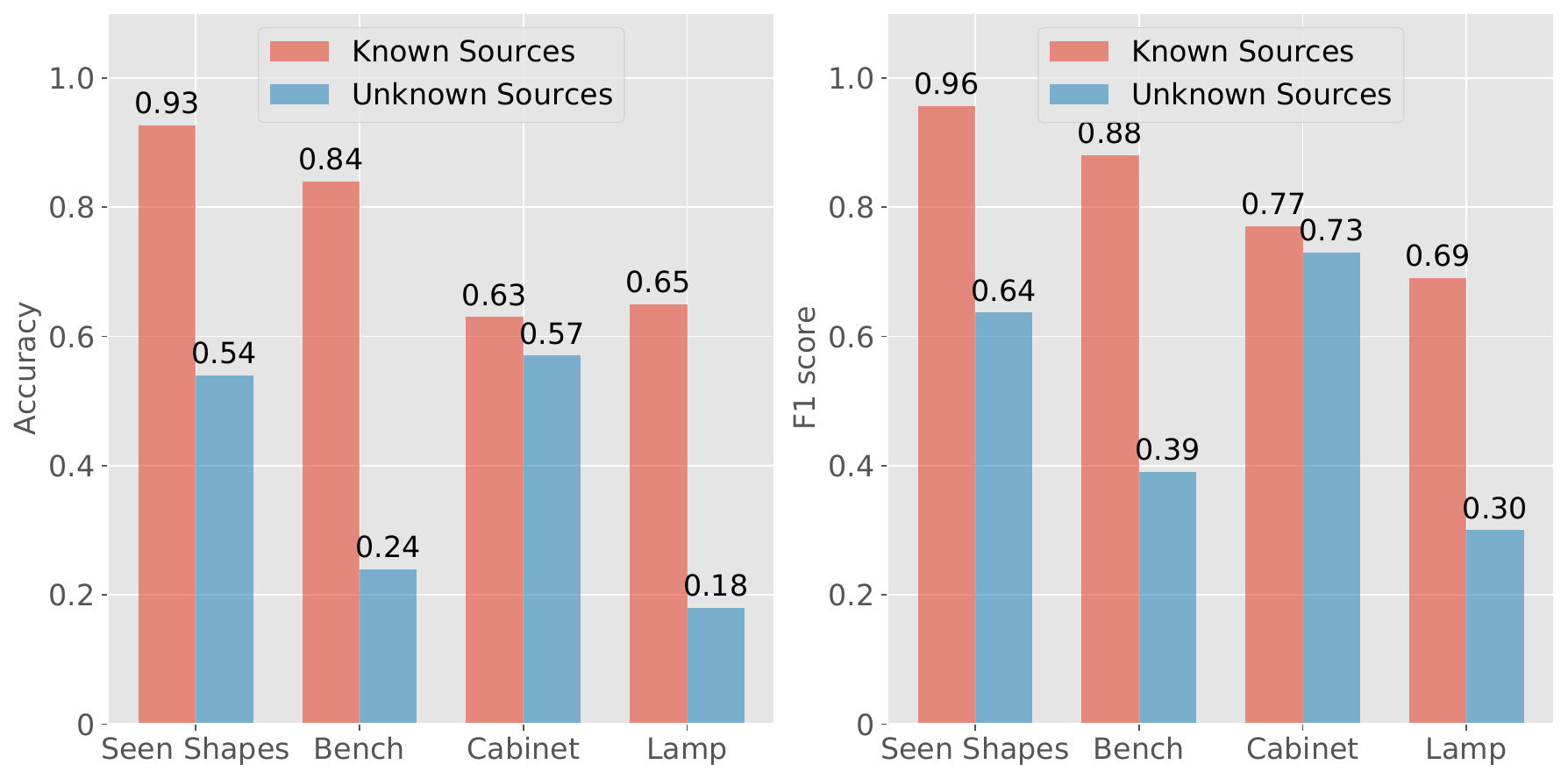}
\caption{\method-PointNet}
\end{subfigure}
\begin{subfigure}{0.85\columnwidth}
\includegraphics[width=\columnwidth]{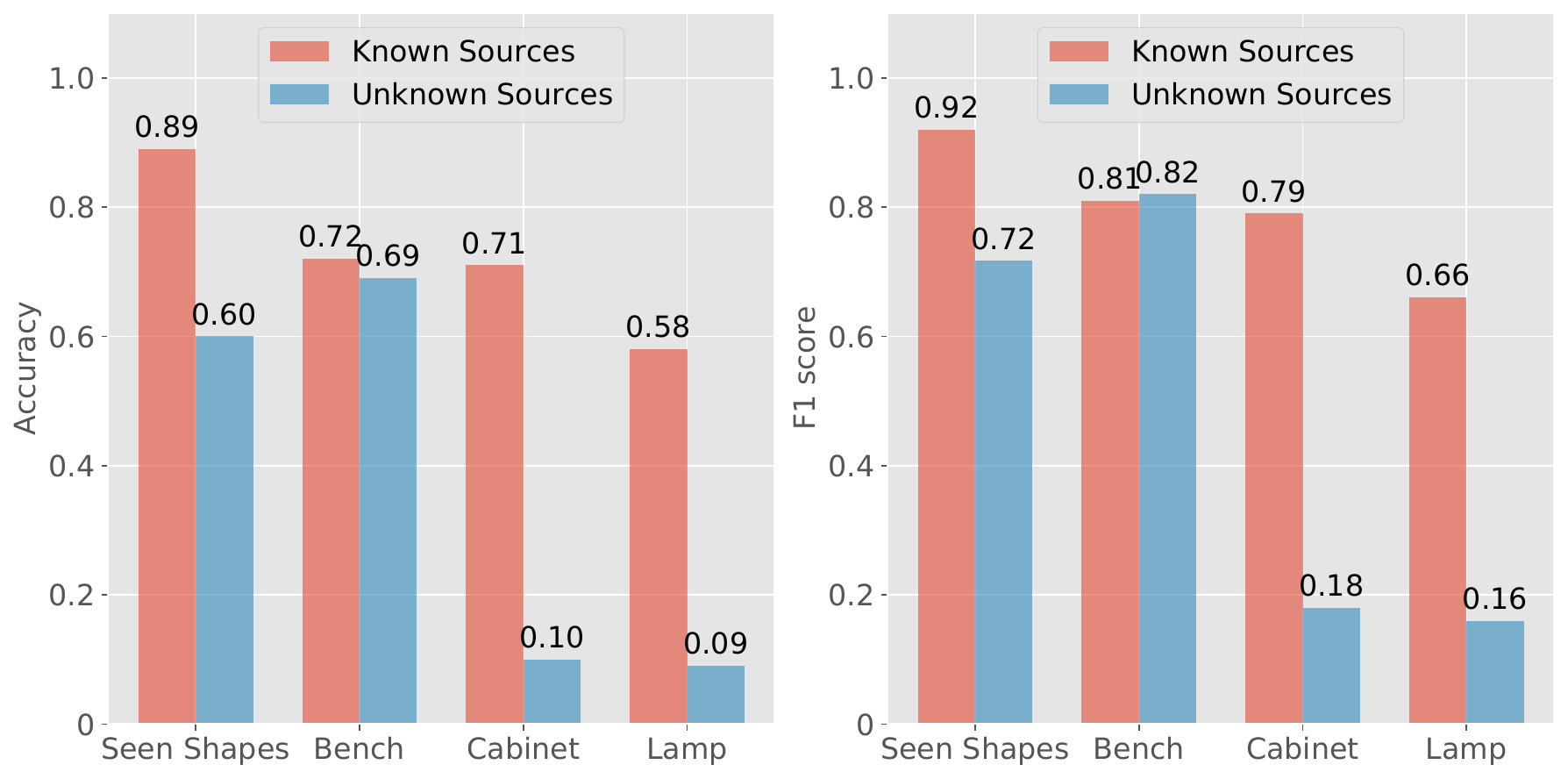}
\caption{\method-DGCNN}
\end{subfigure}
\caption{Attribution performance in the open-world scenario with unknown sources and unseen shapes. 
We report the accuracy and F1 score to evaluate \method-PointNet and \method-DGCNN respectively.
``Seen Shapes'' shows the average accuracy of Airplane, Car, and Chair.}
\label{figure:shape_extended_open_world}
\end{figure*}

\begin{figure}[t]
\centering
\includegraphics[width=0.5\columnwidth]{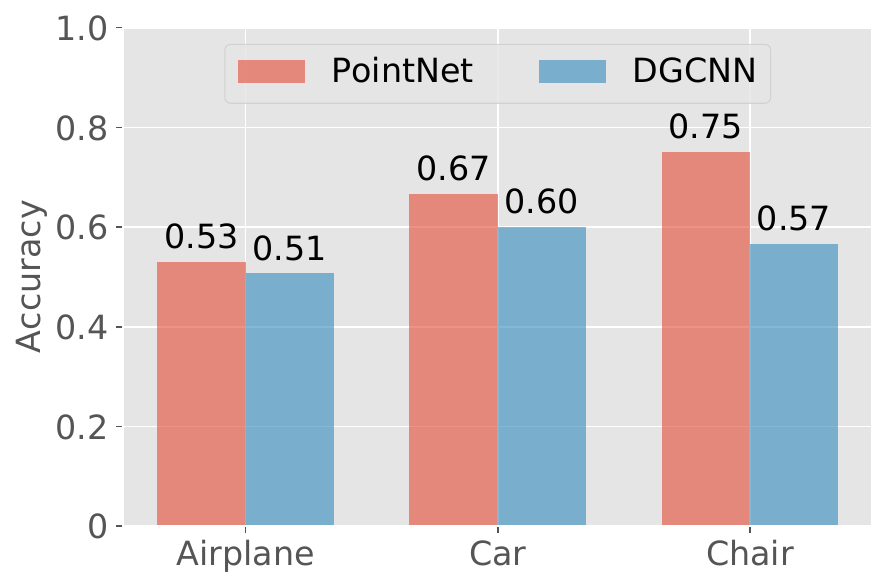}
\caption{Accuracy of \method in differentiating point clouds from two unknown sources: SetVae and PDGN.}
\label{figure:2open_worlds}
\end{figure}

\mypara{Baselines} 
Due to the absence of previous work in synthetic point cloud attribution (especially in \openworld), we compare our framework with several baselines that potentially work.

\begin{itemize}
\item \textbf{Fully-Supervised Training Approach}
We train the PointNet encoder and DGCNN encoder in a supervised manner and leverage the logit probability for attribution.
The models are trained for 200 epochs with a batch size of 32.
We use Adam optimizer with a learning rate of 0.001.
In the threshold assignment stage, we select the lowest logit probability of the training data in the ground-truth class, e.g., 0.95, and use it as a criterion to identify the sources of point clouds.

\item \textbf{Projection-Based Approach}
We project 3D point clouds onto 2D images and utilize a ResNet50 model to extract visual features.
For each point cloud, we obtain three projections and concatenate their respective image features into one final feature, which is then fed into the final classification layer.
The ResNet50 model is trained for 200 epochs with a batch size of 128, using Adam as the optimizer and a learning rate of 0.01.
Regarding threshold assignment, we adopt the same selection strategy as that used in the fully-supervised training approach described above.

\item \textbf{Siamese Training Approach}
As one type of contrastive learning paradigm, a Siamese network learns to differentiate pairs of inputs~\cite{CH21}.
A Siamese network consists of two identical encoders with the sharing parameters.
It takes in a pair of point clouds as input and outputs the logit probability that these originate from the same source.
In every pair of point clouds, there is an equal probability of 50\% of originating from the same source and 50\% from different sources.
We train the Siamese network for 200 epochs with a batch size of 20.
We use the Adam optimizer, a learning rate of 0.001, and a weight decay of 1e-4.
In the attribution phase, we randomly select a pair of point clouds from a known source and a new source respectively, and the model asserts if they originate from the same source.
\end{itemize}

\mypara{Single-Shape Results}
\autoref{table:open_world} shows the comparison result of baselines and \method across three shapes.
\method outperforms all baselines and consistently achieves high accuracy (0.73-1.00) and F1 score (0.85-1.00) when recognizing point clouds from unknown sources.
For known sources, \method achieves an accuracy between 0.82 and 0.96 and an F1 score ranging from 0.89 to 0.98.
Take Car as an example, \method-D achieves 0.98 and 1.00 F1 score on known and unknown sources respectively, while the best accuracy score from all baselines is 0.96 and 0.90.
The best-performing baseline is the Siamese training approach, which shows a wide accuracy range from 0.14 to 1.00, occasionally matching \method in certain shapes, such as Chair.
For fully supervised and project-based baselines, we observe good attribution performance for known sources, however, they often fail to identify point clouds from unknown sources.
This result suggests that contrastive learning (supervised contrastive learning in \method and the Siamese baseline) has a better performance in identifying open-world examples compared to the supervised training paradigm (fully supervised and projection-based baselines).

In the above evaluation, we use the optimal thresholds determined in \autoref{subsection:experimental_setup}.
They are determined with the respective validation set, which has the same distribution as the testing examples.
In the real world, this setting is often not realistic, as the validation set is not always available.
To evaluate whether \method can generalize to more unknown sources, we introduce two more generative models and test \method's performance on these new examples.
The result reveals that \method can still recognize point clouds from new unknown sources with an F1 score between 0.71-0.90.
We present the details in \autoref{appendix: generalizability_unknown_sources}.

\mypara{Multiple-Shape Results}
Here, we evaluate the most challenging task: attributing unseen shapes to unknown sources.
This scenario inevitably impacts attribution accuracy due to two-fold distractions: the unknown sources and the divergence in shape distribution caused by unseen shapes.
The results are depicted in \autoref{figure:shape_extended_open_world}.
Although \method fails to attribute unseen point clouds to unknown sources, it manages to attain decent performance for known sources (0.58-0.84 accuracy).
For example, when attributing the unseen shape Bench, which resembles the shape of a Chair, the accuracy reaches 0.84 with \method-PointNet and 0.72 with \method-DGCNN for known sources.

\mypara{Multiple Unknown Sources Differentiation}
In the above evaluation, \method classifies point clouds from all unknown sources into a single class, i.e., Unknown.
We then investigate whether \method can differentiate various unknown sources (PDGN and SetVae in our evaluation).
To this end, we employ Gaussian clustering on the point cloud feature vectors from these two unknown sources.
Specifically, we input these feature vectors into a Gaussian mixture model with two components to obtain the predicted clusters.
From these clusters, we derive the classification results for the two sources.
\autoref{figure:2open_worlds} presents the differentiation results of three observed shapes (Airplane, Car, and Chair) from different unknown sources.
The differentiation accuracy scores exceed the random guess baseline (0.5) for all three shapes.
For example, \method-PointNet achieves an accuracy of 0.75 in separating Chair features between PDGN and SetVae.
This result demonstrates that \method may further capture the differences among unknown sources.

\mypara{Visual Analysis}
We use t-SNE to project the point cloud features obtained from \method to a 2-dimensional space.
The visualizations are shown in \autoref{figure:tsne_visualize_pointnet} and \autoref{figure:tsne_visualize_dgcnn} in the Appendix.
Among all three shapes, Car and Chair demonstrate satisfactory clustering performances, where embedding projections from different sources are distinguishable from each other.
We notice that it is difficult to separate Airplane point clouds between PointFlow and SetVae.
Nevertheless, the projections from two unknown sources are separated for all three shapes, indicating that \method can recognize more than one unknown source.

\mypara{Takeaways}
Based on the above analysis, we have several takeaways: 1) \method can reliably attribute seen shapes to both known and unknown sources;
2) It can further distinguish point clouds from multiple unknown sources;
3) The attribution ability can generalize to point clouds of unseen shapes, in particular, those unseen shapes that are visually close to the seen ones.

\begin{figure*}
\centering
\begin{subfigure}{1.1\columnwidth}
{\includegraphics[width=\columnwidth]{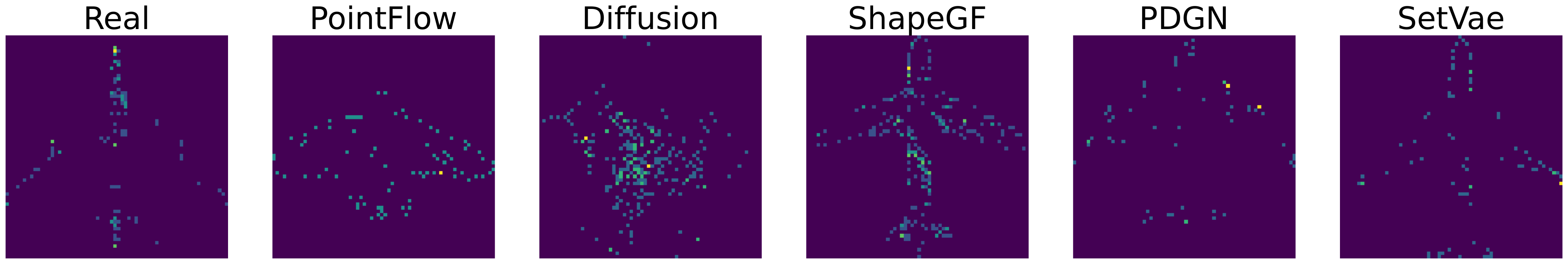}}
\end{subfigure}
\begin{subfigure}{1.1\columnwidth}
{\includegraphics[width=\columnwidth]{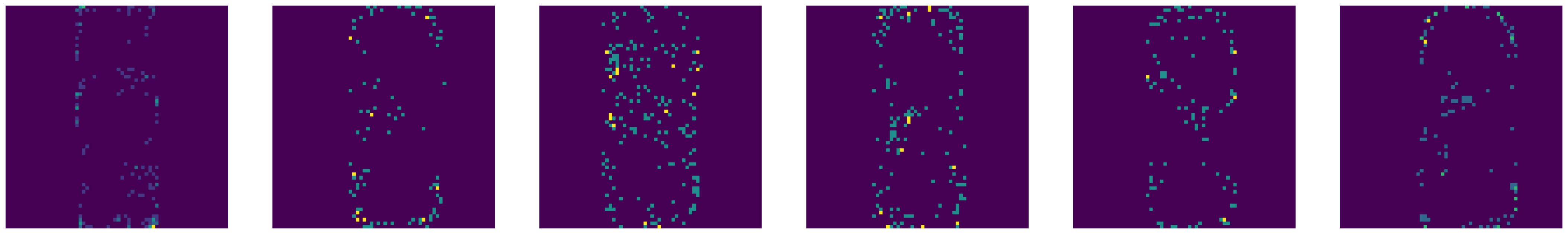}}
\end{subfigure}
\begin{subfigure}{1.1\columnwidth}
{\includegraphics[width=\columnwidth]{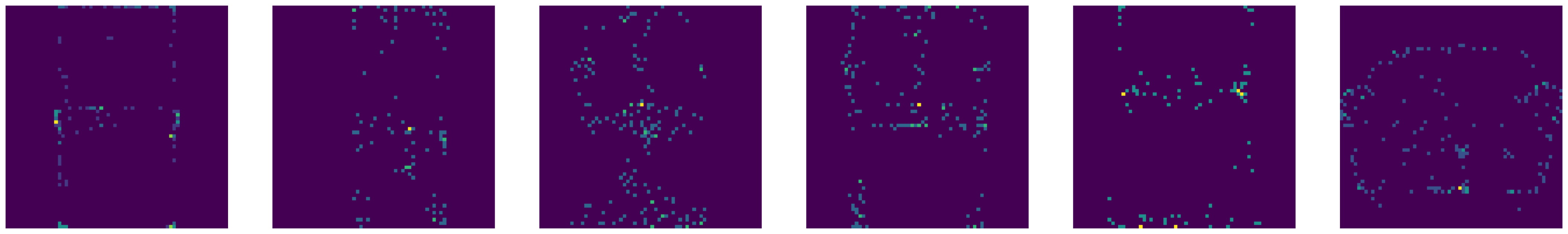}}
\end{subfigure}
\caption{Visualization of six fingerprints on three shapes: Airplane, Car, and Chair, from top to bottom. 
These fingerprints are built by stacking 100 depth images projected from critical points from each source.}
\label{figure:fingerprint}
\end{figure*}

\section{Explainable Attribution}
\label{section: explainable_attribution}

In this section, we discuss the nature of source attribution and introduce an approach to visualize the unique patterns in point clouds associated with each source.
This explains how \method recognizes point clouds from different sources.

\subsection{Behind the Attribution}

Generative models synthesize fake point clouds by sampling from a distribution that is learned using real-world data.
The sampled distribution, in practice, is often biased toward the real-world data distribution.
Take Generative Adversarial Neural Network (GAN) as an example.
Due to the imperfect convergence during the training process, the generator and the discriminator often fall into the sub-optimal solution, such that GAN only partially describes the data distribution from the real world.
The bias between the sampled distribution and real distribution has made the source attribution possible.
The attribution model captures the unique bias by transforming point clouds into feature vectors and learning to differentiate them in a supervised or self-supervised fashion.

\mypara{Critical Points}
To explain how the attribution model differentiates point cloud feature vectors, we introduce \emph{critical points}~\cite{QSMG17}, which is defined as a set of points that significantly affect the feature vectors.
Critical points contain the most geometric and semantic information about point clouds.
We identify critical points of a point cloud by tracing the points that contribute the most to the maximum pooled feature obtained from PointNet.
To ensure permutation invariance, which states that the order of points is invariant to the feature extraction, PointNet calculates the individual point feature or local geometric feature and then aggregates the local features into a global descriptor with a symmetric max-pooling function.
For instance, PointNet transforms a point cloud of dimension (2048, 3) into a feature map of dimension (2048, 1024) before down-sampling into a global feature whose length is 1024 by max-pooling.
We trace and record the points of these largest values in the feature map (input of the max-pooling layer) as critical points.
Examples of critical points are demonstrated in \autoref{figure:critical_points_airplane} in the Appendix.
In this figure, we present three columns of point clouds, each representing a different type of airplane.
Comparing the critical points within one column, we can see how \method identified different points/areas in point clouds from different sources.
More details on selecting these examples are presented in \autoref{appendix: critical_points_visualization}.

\begin{figure*}[!t]
\centering
\includegraphics[width=0.55\textwidth]{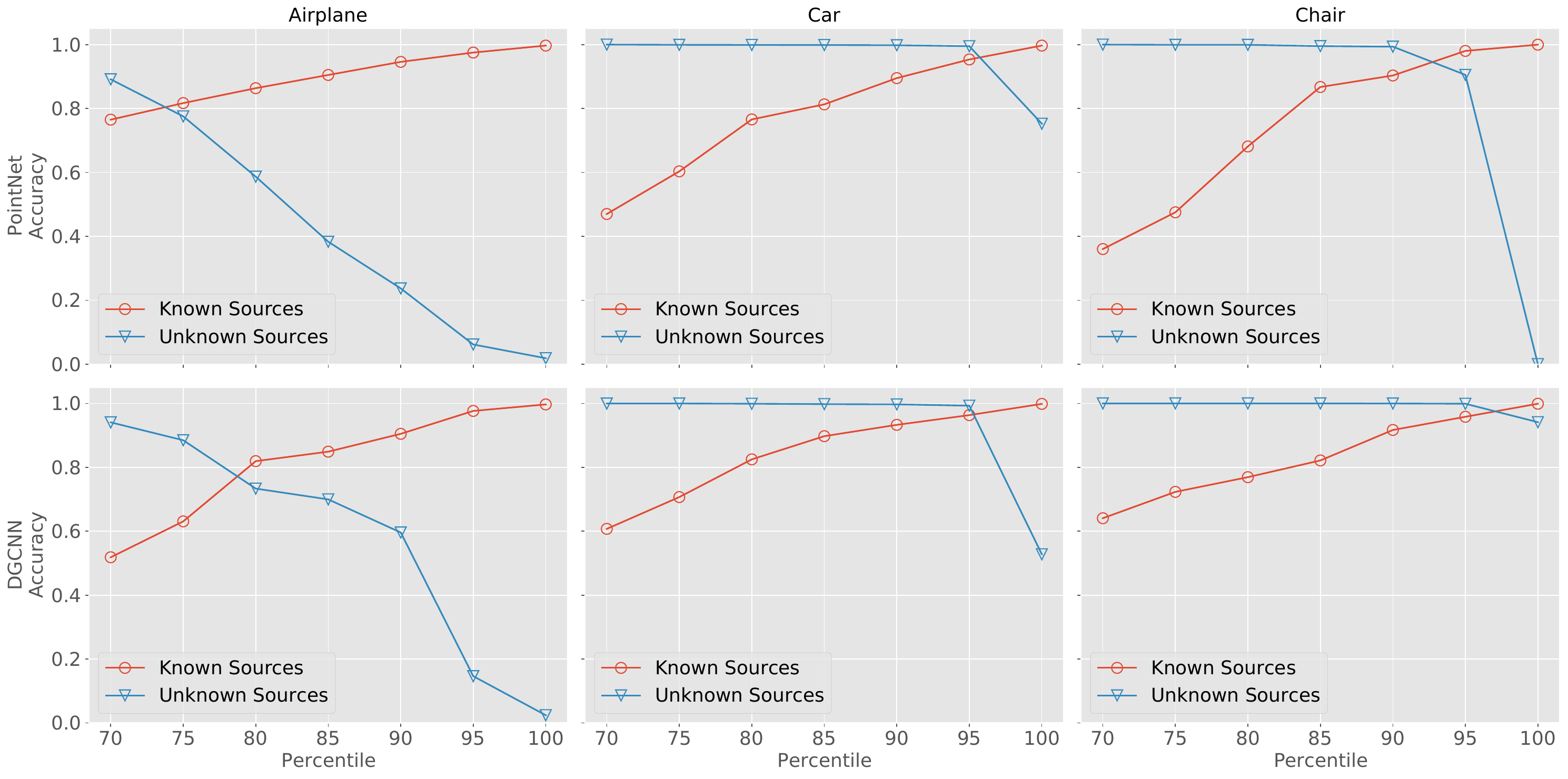}
\caption{Attribution performances on known and unknown sources when the threshold is increasing. 
From left to right, we display the attribution results of Airplane, Car, and Chair.}
\label{figure:threshold}
\end{figure*}

\subsection{Fingerprint Visualization}

Fingerprints are persistent and identifiable patterns that could be distinguished from other sources/models.
There are multiple ways to build fingerprints of image generative models~\cite{flairNLP, YDF19}.
However, to our knowledge, none of the existing work explores the fingerprint associated with point cloud models.
Here, we introduce how to visualize the fingerprints of the point cloud generative models and verify the uniqueness of the fingerprints qualitatively.

To build fingerprints, we first extract the critical points during the attribution process.
Specifically, we randomly select 100 sets of critical points of a generative model (or collected from the real world) of the same shape, e.g., 100 airplanes.
We then project the critical points to the same plane, e.g., x-y plane, and obtain 100 depth images.
By stacking these depth images to form an average depth image that shows the frequency spectra, we extract and visualize the unique patterns of each source.

Fingerprints of six different sources on three shapes are visualized in \autoref{figure:fingerprint}.
For the same shape shown in each row, every source presents a unique pattern that distinguishes it from other sources.
Take Airplane from the Real source, as an example, points at the airplane's head, end of the wings, and tail have higher frequencies than other areas.
An airplane whose critical points set is similar to this pattern is likely to be attributed to the Real source.
Similarly, for airplanes generated by ShapeGF, the critical points mainly describe the airplane's skeleton.
Overall, the above analysis enhances the understanding of how \method conducts the source attribution task.

\section{Ablation Study}
\label{section:ablation_study}

In this section, we conduct a series of ablation studies to investigate how attribution performance can be affected by several factors.
We start with investigating how threshold selection in \method influences the performances of both \closeworld and \openworld.
Second, we explore whether the feature vector dimension, i.e., the output dimension of encoding backbones, affects the attribution performance.
In addition, we study the effect of the close-world pre-training stage in attributing point clouds compared to performing open-world pre-training from scratch (without the close-world pre-training stage).
Finally, we evaluate the robustness of \method against common perturbations on the point clouds.

\begin{figure*}[t]
\centering
\includegraphics[width=0.5\textwidth]{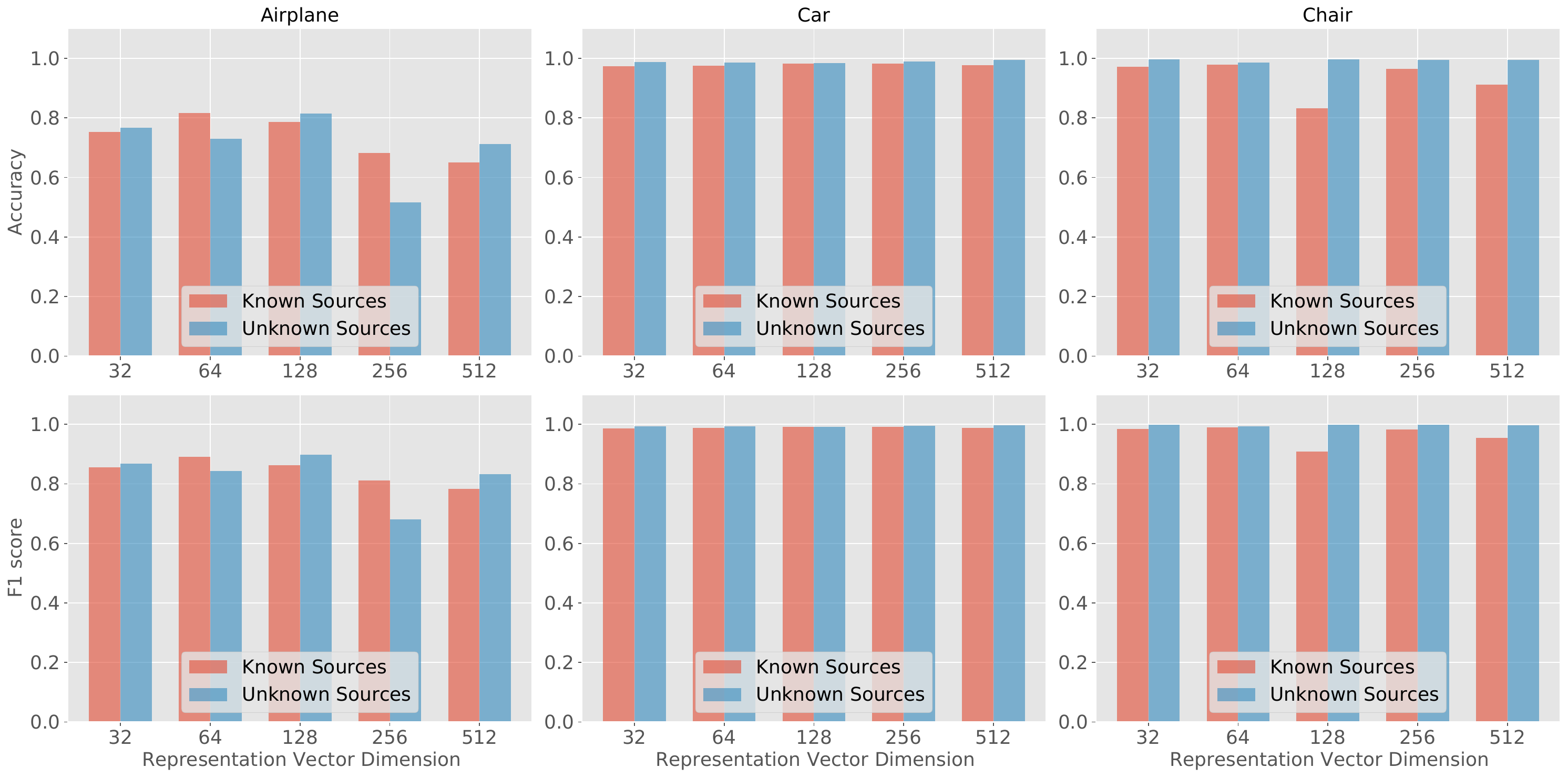}
\caption{Attribution performances on both known and unknown sources when feature vector dimension is varied.
Attribution performances of Airplane on both worlds are more balanced when the dimension is 128.}
\label{figure:dimension}
\end{figure*}

\begin{figure*}[t]
\centering
\includegraphics[width=0.5\textwidth]{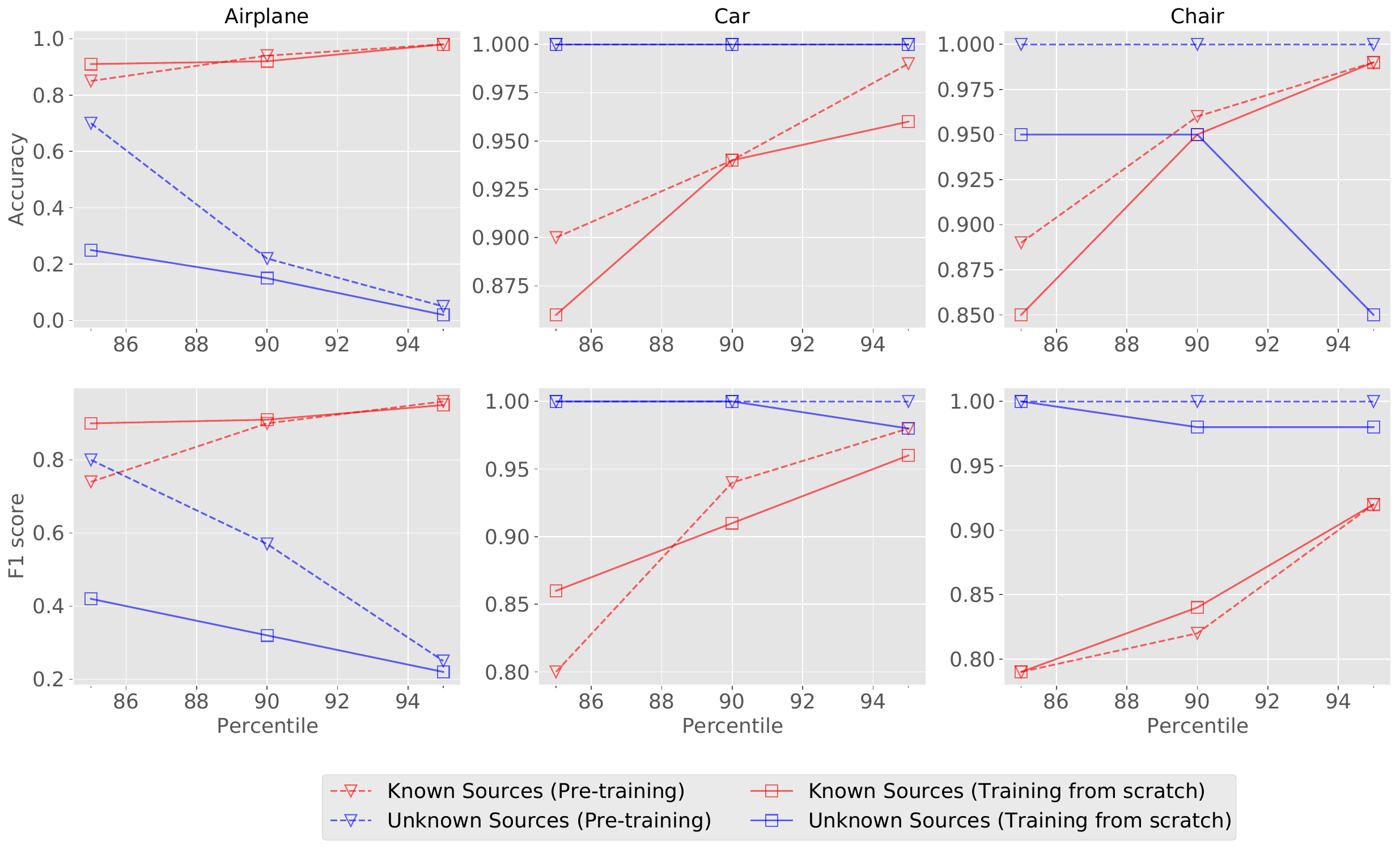}
\caption{Attribution performances when the encoder is pre-trained and trained from scratch. 
Adopting a pre-trained encoder leads to better attribution accuracy when recognizing unknown sources.}
\label{figure:pretrain}
\end{figure*}

\mypara{Effect of Threshold Selection}
According to our threshold-based assignment principle, threshold selection can affect the attribution accuracy of both known sources and unknown sources.
\autoref{figure:threshold} shows the trend of attribution accuracy in both settings when we enlarge the threshold by increasing the $P$-percentile gradually.
As the threshold increases, the attribution accuracy of unknown sources demonstrates an evident rising trend, as the testing examples are more inclined to be assigned to known sources.
Meanwhile, the accuracy of identifying known sources exhibits a reversed or steady trend depending on different shapes.
For the attributing result of Airplane in the first column, we capture a trade-off relation with the intersection point at the range of 75\%-80\%.
The accuracy on unknown sources maintains a high level for other shapes that can be differentiated more easily, and intersection appears after we set a higher percentile.

\begin{figure*}[t]
\centering
\includegraphics[width=0.75\textwidth]{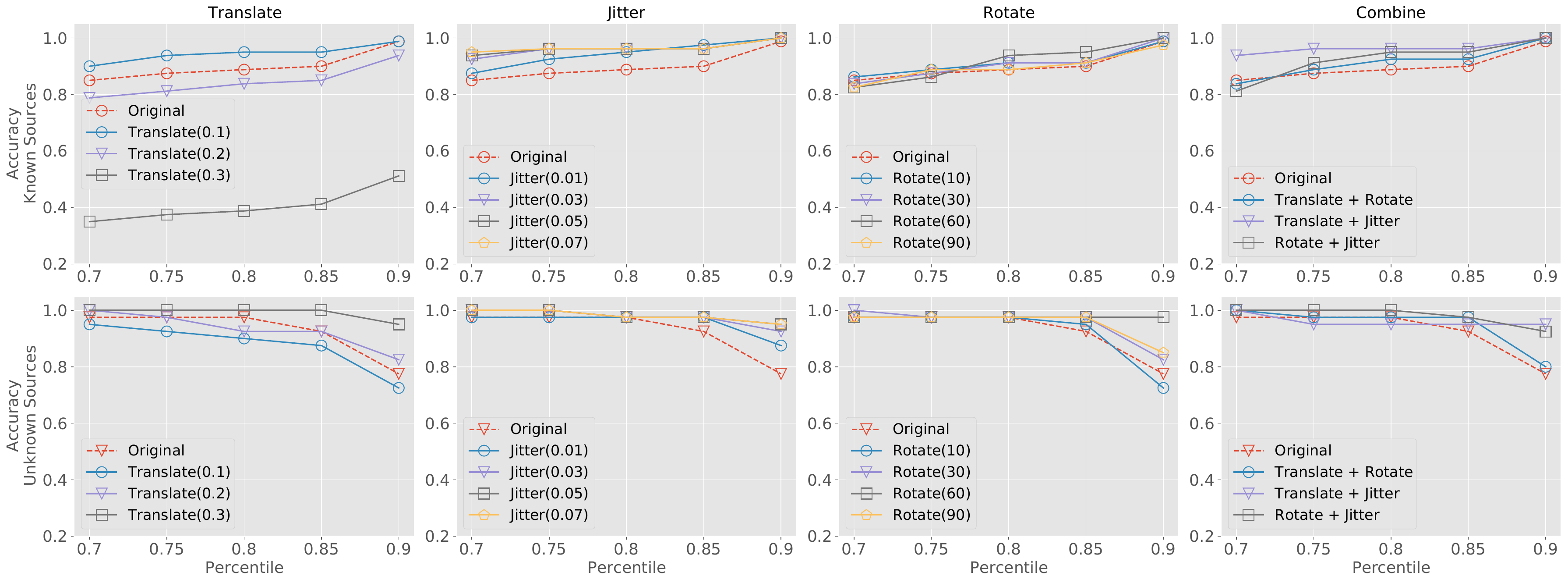}
\caption{Attribution performances on Car against perturbations. 
Perturbations include \textit{translate}, \textit{jitter}, \textit{rotate}, and the three combined.
The attribution accuracy on both known and unknown sources is robust against most of the applied perturbations.}
\label{figure:perturbation}
\end{figure*}

\mypara{Effect of Feature Vector Dimension}
We vary the dimension from 32 to 512 and evaluate the performance of \method on Airplane, Car, and Chair.
\autoref{figure:dimension} demonstrates the attribution performances on varying dimensions.
For shapes such as Car and Chair, various dimensions do not significantly affect the attribution performance to a large extent.
However, for point clouds that are relatively hard to differentiate, e.g., Airplane, we obtain the most balanced attribution performance when the dimension size is 128, under the same threshold setting.
Note that we adopt 128 as the feature vector dimension in the main experiments, following the work\cite{KTWSTIMLK20}.

\mypara{Effect of Close-World Pre-training Stage}
To study how the encoder pre-training stage in \method affects the \openworld, we compare the attribution performances when the encoder is randomly initiated or initiated with the parameters updated from the pre-training stage.
\autoref{figure:pretrain} displays the comparison results of attributing three shapes, i.e., Airplane, Car, and Chair.
We vary the $P$-percentile from 85 to 95 to gradually increase the threshold values.
\autoref{figure:pretrain} shows that \method with the pre-training encoder performs almost equally as well as the training-from-scratch encoder when attributing data to known sources.
However, the pre-training stage benefits the attribution accuracy of unknown sources to a certain extent.

\mypara{Effect of Perturbations}
We study the effect of common perturbations on point clouds to evaluate the robustness of \method.
\autoref{figure:perturbation} shows the result after we exert different perturbations on the testing examples.
We observe that \textit{jitter}, \textit{rotate}, and combined perturbations do not degrade attribution accuracy both on known and unknown sources, which demonstrates the resistance of \method to these perturbations.
However, we observe a decline when we translate point clouds in different coordinate frames, especially when the coordinate origin is distant from the old origin.
The conjectured reason is that the point cloud features are not aligned in the feature space when point clouds are in different coordinate frames, resulting in a biased distance distribution for threshold-based attribution.

\section{Related Work}
\label{section:related_work}

\mypara{Point Cloud Generation}
Generative models of point clouds can be broadly classified into three categories: Generative adversarial networks (GAN) based models, flow-based models, and variational auto-encoders (VAE) based models.
To adapt GAN~\cite{GPMXWOCB14} in the point cloud domain, PC-GAN~\cite{LZZPS18} modifies GAN model to learn a hierarchical sampling process for point cloud generation.
As an end-to-end generation model, PDGN~\cite{HXXQY20} proposes a progressive GAN network composed of multiple stacking deconvolution networks.
Flow-based models~\cite{SZZC19} explicitly model the density function using continuous normalizing flow~\cite{RM15}.
PointFlow~\cite{YHHLBH19} generates 3D point clouds by learning a distribution of distributions and models each distribution as an invertible parameterized transformation of 3D points from a prior distribution.
Other novel works such as ShapeGF~\cite{CYAHBSH20} model a shape by learning the gradient field of its log density, and then moving the points gradually from a generic prior distribution towards the surface. Diffusion models~\cite{LH21} generate point clouds in the perspective of thermodynamics.

\mypara{Point Cloud Classification}
Different from 2D images, 3D point cloud classification and segmentation tasks cannot be achieved with standard deep neural network models (i.e., MLP) due to irregular structure.
PointNet~\cite{QSMG17} is the first work to directly forward raw point cloud data to the DNN model and obtain global features leveraging the symmetric max pooling function.
Inspired by PointNet and convolution modules, follow-up works such as A-CNN~\cite{KZH19} PointGrid~\cite{LD18}, ShellNet~\cite{ZHY19}, and PointCNN~\cite{LBSWDC18} learn the local features by adapting convolution modules to point clouds.
Graph neural networks~\cite{SGTHM09} are also applied in learning point cloud features.
Wang et al. \cite{WSLSBS19} propose DGCNN by constructing local graphs and learning the edge information that describes the relationships between a point and its neighbors.

\mypara{Deepfake Image Detection and Attribution}
Deepfake detection and attribution already emerged in 2D image~\cite{YSCDF20,YSAF21,YDF19,LQT20,ZZCWZY21} and text domain~\cite{GZR20,PEZZRL20}.
For 2D images, existing work~\cite{WWZOE20,ZZCWZY21,YDF19} has demonstrated that deepfake image detection could achieve a favorable accuracy in the \closeworld.
Wang et al.\ \cite{WWZOE20} and Zhao et al.\ \cite{ZZCWZY21} detect fake images in the close world by training binary or multi-class classifiers.
Ning et al.\ \cite{YDF19} attribute fake images to its GAN model by learning and analyzing image fingerprints.
Previous studies~\cite{GSRS21,YSAF21,YSCDF20} on tracing data in the open world also received much attention.
Ning et al.\ \cite{YSAF21,YSCDF20} actively embed fingerprints into generative models, such that open-world models can be verified by decoding and matching fingerprints.
Recently, Sha et al.\ \cite{SLYZ23} carry out a systematic study on the detection and attribution of fake images generated by the latest text-to-image generation models, including stable diffusion, DALLE-2, GLIDE, etc.
Inspired by the success of deepfake image detection and attribution, we explore the possibility of detecting fake point clouds in this work.

\section{Conclusion}
\label{section:conclusion}

In this paper, we proposed an attribution framework \method to pioneer the study of point cloud authentication and source attribution.
This framework can attribute point clouds to both known and unknown sources.
We also study the generalizability of attribution on different shapes by training \PerShape models and \ShapeExtended models.
Evaluation results verified the effectiveness of \method when attributing both known and unknown sources, even shapes that are not included in the training set.
The fingerprint visualization we introduced assists in understanding how our framework recognizes the critical points in point clouds and captures the uniqueness of each generative model.
We hope that our study can also foster research in driving responsible and trustworthy artificial intelligence in point cloud generative models to prevent them from mischievous use.

\mypara{Limitations and Future Work}
This work has limitations.
The first key area for improvement is to expand the number of generative models.
In the main experiments, we currently only include four known sources and two unknown sources.
In future work, we will explore incremental learning to incrementally involve more generative models on top of the trained attribution model.
Second, when evaluating the performance of \method, we rely on a small number of samples in the validation set to determine the optimal threshold, which is often not realistic in practice.
We test two additional unknown sources in the Appendix and found that the threshold remains effective, although in a limited capacity, on more unknown sources.
We will continue exploring the best approach to determine thresholds for recognizing closed-world and open-world examples.

\mypara{Acknowledgement}
We thank all reviewers for their constructive suggestions.
This work is partially funded by the European Health and Digital Executive Agency (HADEA) within the project ``Understanding the individual host response against Hepatitis D Virus to develop a personalized approach for the management of hepatitis D''(D-Solve) (grant agreement number 101057917).

\begin{small}
\bibliographystyle{plain}
\bibliography{normal_generated_py3}
\end{small}

\appendix

\section{Generalizability to More Unknown Sources}
\label{appendix: generalizability_unknown_sources}

In the \openworld, \method presents satisfactory performance in identifying point clouds from two unknown sources: PDGN and SetVae.
To evaluate whether \method can generalize to more unknown sources, especially when their validation sets are not accessible,  we introduce two additional generative models, GPointNet~\cite{xie2021generative} and SoftFlow~\cite{kim2020softflow}.
For each model, we generate 1,000 point clouds of the shape Airplane, Car, and Chair, respectively.
We directly adopt the optimal thresholds obtained in \autoref{table:optimal_thresholds} and measure \method's performances in attributing these samples.
\autoref{table:additional_unknown_sources} displays the attribution performance.
\method identifies the point clouds from unknown sources with an accuracy ranging from 0.41 to 0.81 across three shapes.
The best performance is observed in Car, where \method-P achieves an accuracy of 0.81 and an F1 score of 0.90.
This finding suggests that \method may generalize to more unknown sources with previously determined thresholds in \autoref{table:optimal_thresholds}.

\begin{table}[!ht]
\centering
\caption{Attribution performance of \method in identifying point clouds from additional unknown sources: GPointNet~\cite{xie2021generative} and Softflow~\cite{kim2020softflow}.}
\label{table:additional_unknown_sources}
\scalebox{0.8}{
\begin{tabular}{lcccccc}
\toprule
&\multicolumn{2}{c}{Airplane}& \multicolumn{2}{c}{Car} & \multicolumn{2}{c}{Chair} \\
& Acc & F1 score & Acc & F1 score & Acc & F1 score \\
\midrule
\method-P & 0.41 & 0.58 & \textbf{0.55} & \textbf{0.71} & \textbf{0.81} & \textbf{0.90} \\
\method-D & \textbf{0.57} & \textbf{0.73} & 0.51 & 0.68 & 0.43 & 0.60 \\
\bottomrule
\end{tabular}
}
\end{table}

\begin{figure*}[!ht]
\centering
\begin{subfigure}{0.95\textwidth}
\centering
\includegraphics[width=\textwidth]{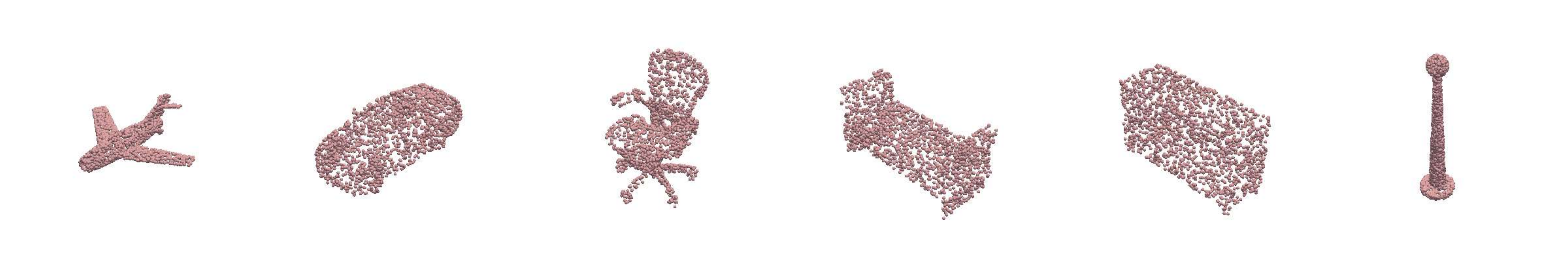}
\caption{Real}
\end{subfigure}
\hfill
\begin{subfigure}{0.95\textwidth}
\centering
\includegraphics[width=\textwidth]{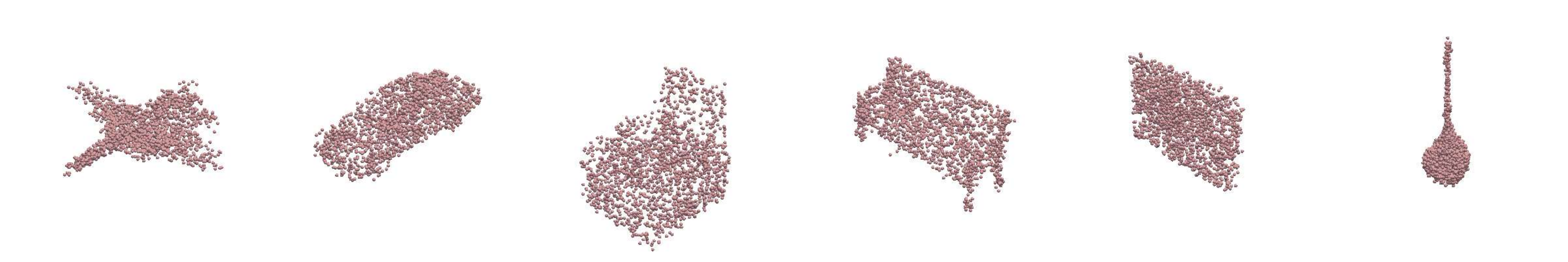}
\caption{PointFlow}
\end{subfigure}
\hfill
\begin{subfigure}{0.95\textwidth}
\centering
\includegraphics[width=\textwidth]{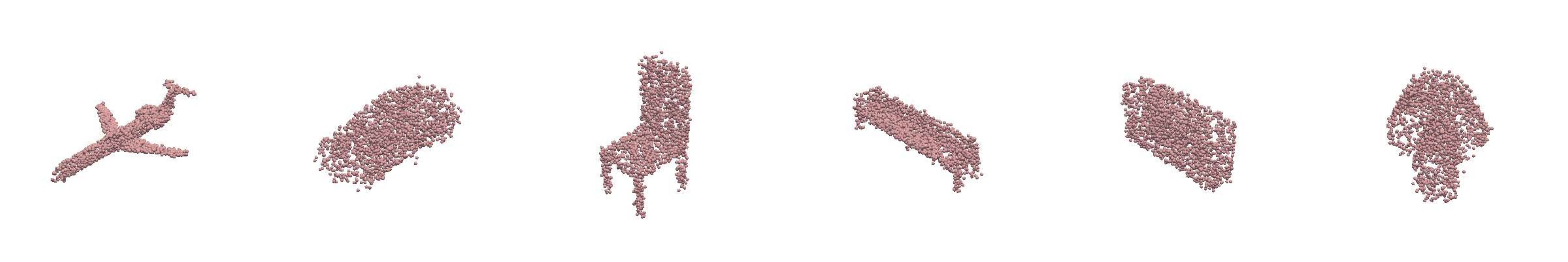}
\caption{Diffusion}
\end{subfigure}
\hfill
\begin{subfigure}{0.95\textwidth}
\centering
\includegraphics[width=\textwidth]{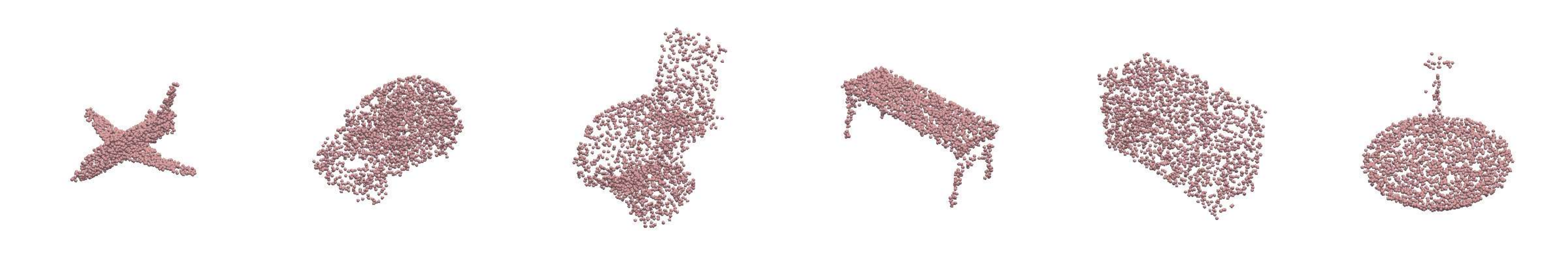}
\caption{ShapeGF}
\end{subfigure}
\hfill
\begin{subfigure}{0.95\textwidth}
\centering
\includegraphics[width=\textwidth]{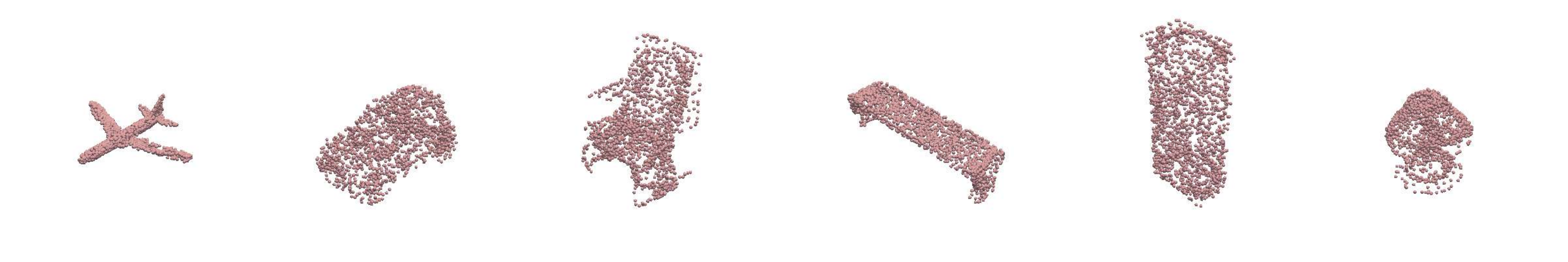}
\caption{PDGN}
\end{subfigure}
\hfill
\begin{subfigure}{0.95\textwidth}
\centering
\includegraphics[width=\textwidth]{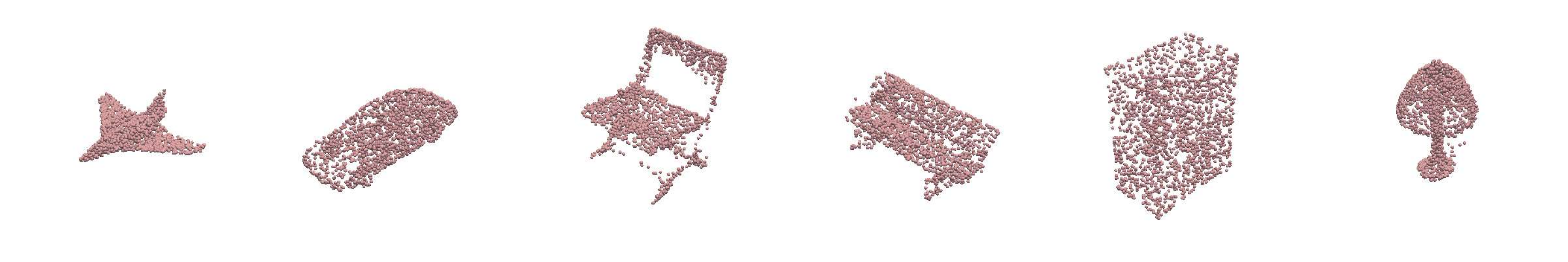}
\caption{SetVae}
\end{subfigure}
\caption{Generated examples from five generative models plus the real world. 
From left to right, we display the shapes: Airplane, Car, Chair, Bench, Cabinet, and Lamp.}
\label{figure:generative_examples}
\end{figure*}

\begin{figure*}[!ht]
\centering
\begin{subfigure}{0.3\textwidth}
\includegraphics[width=\textwidth]{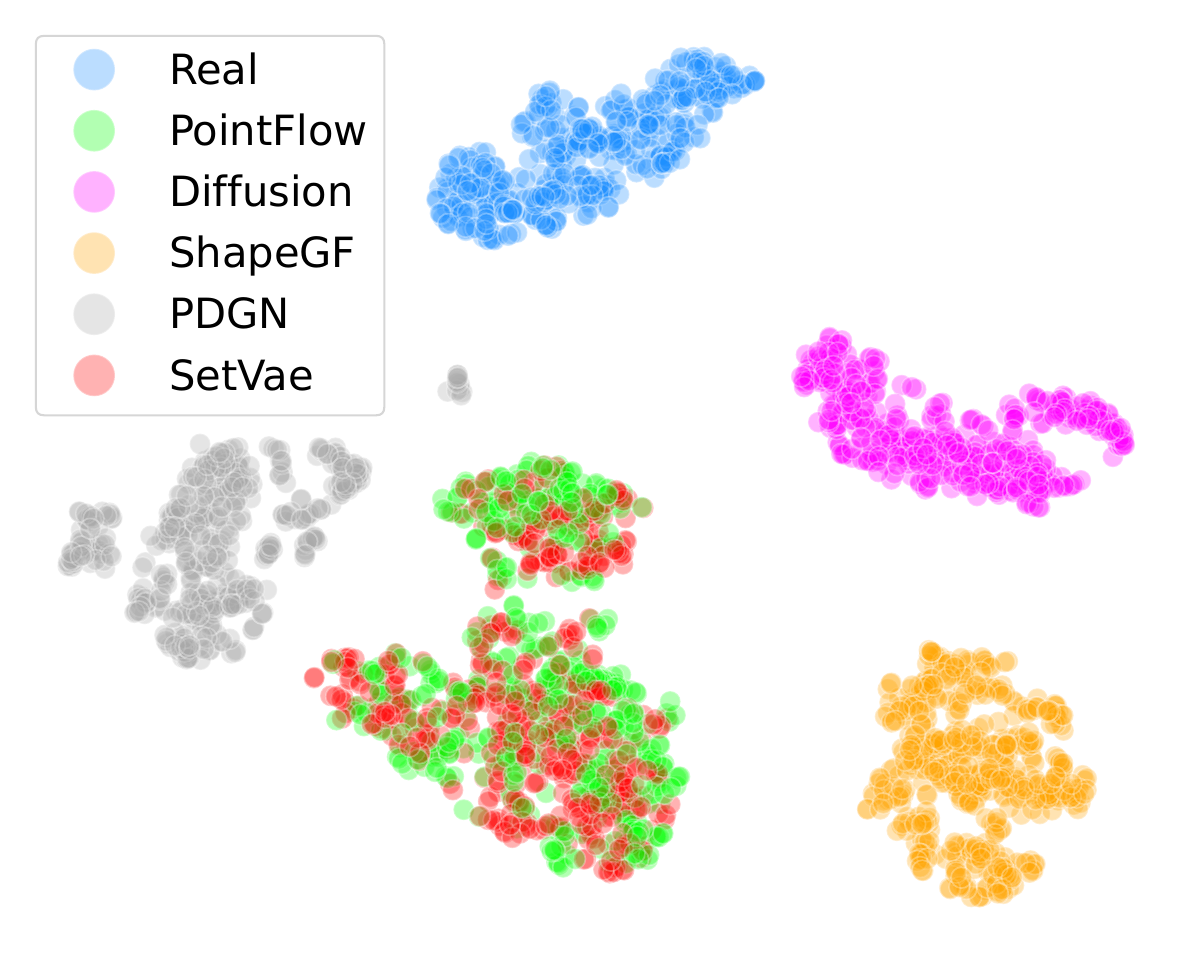}
\caption{Airplane}
\end{subfigure}
\hfill
\begin{subfigure}{0.3\textwidth}
\includegraphics[width=\textwidth]{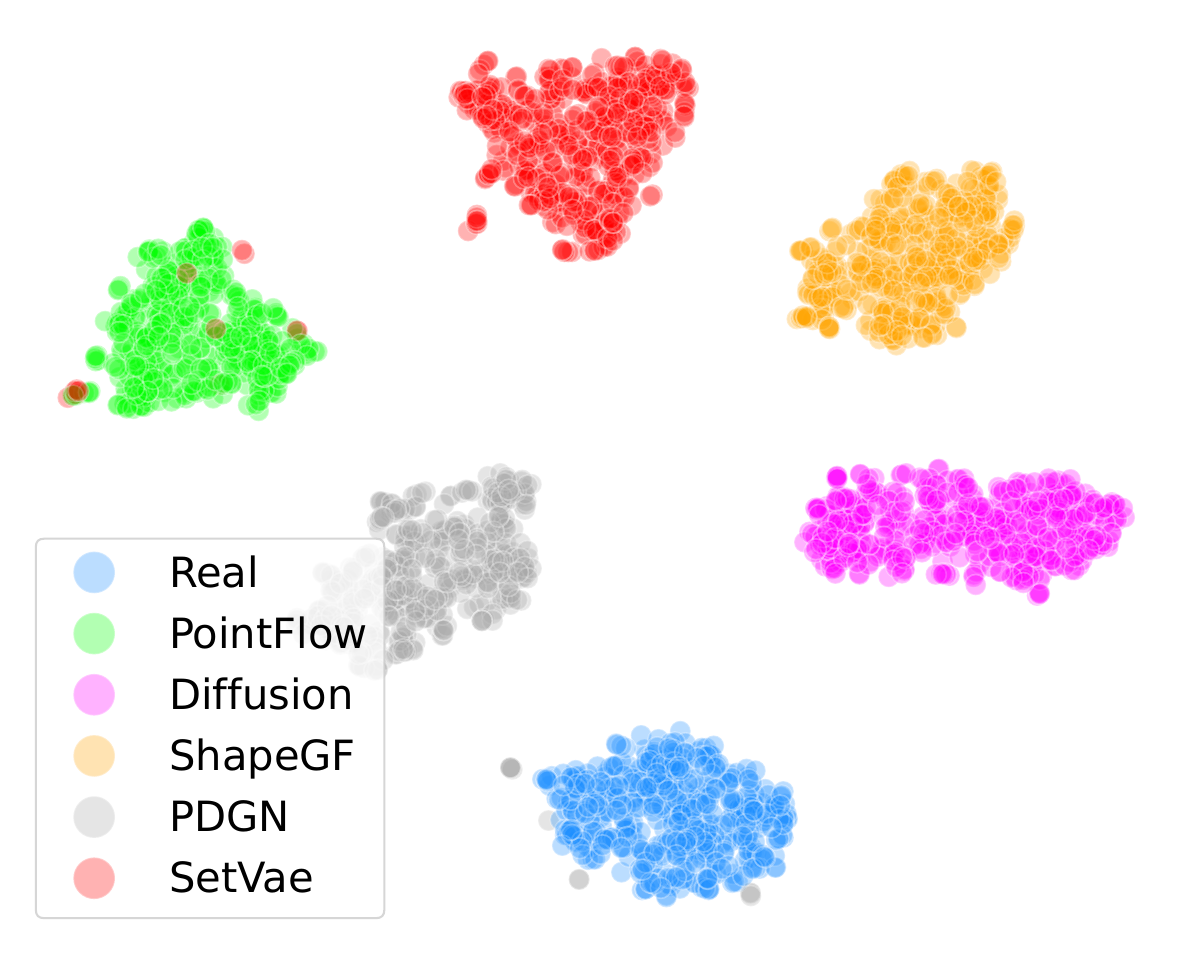}
\caption{Car}
\end{subfigure}
\hfill
\begin{subfigure}{0.3\textwidth}
\includegraphics[width=\textwidth]{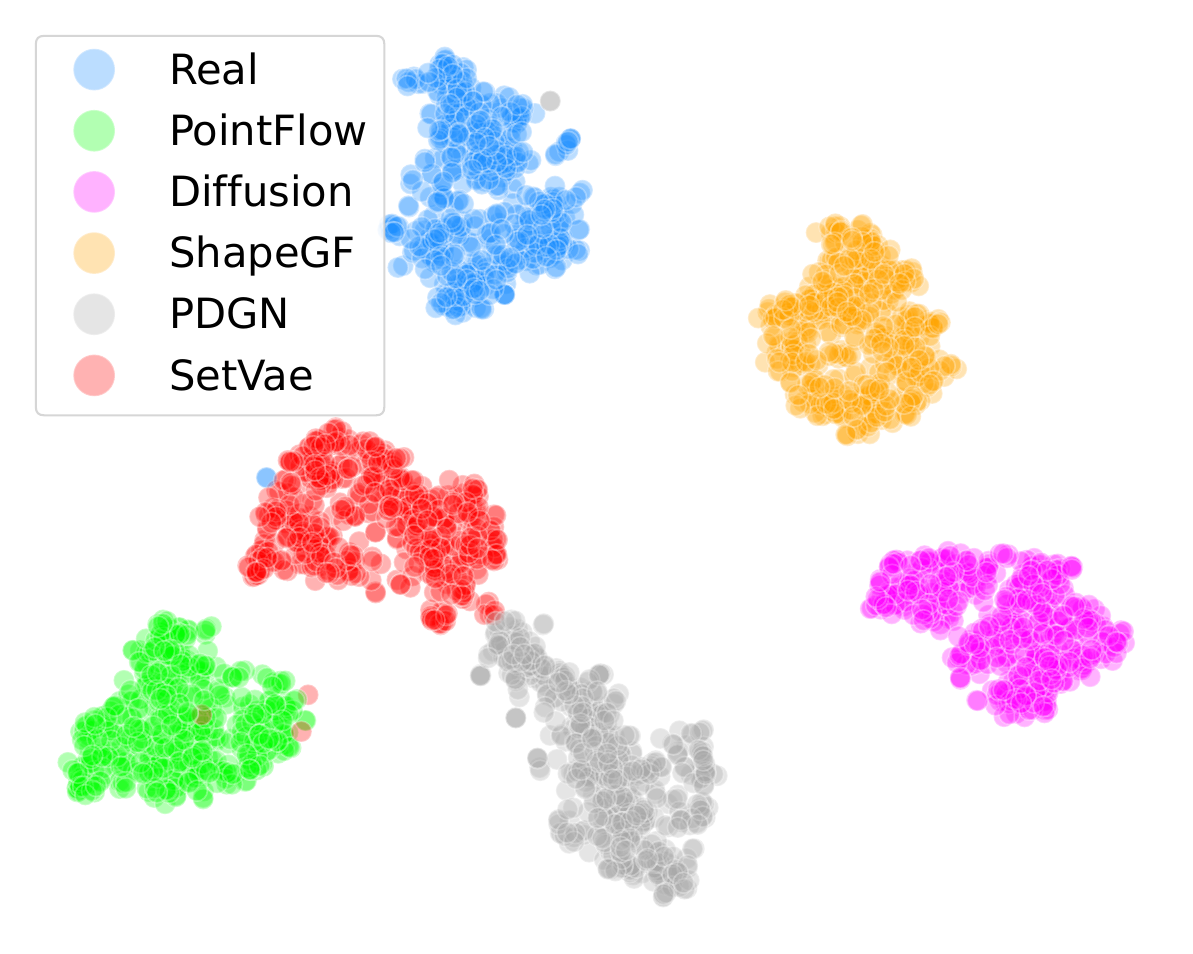}
\caption{Chair}
\end{subfigure}
\caption{t-SNE visualization of clustered point cloud features obtained from \method-PointNet (better viewed in color). 
PDGN and SetVae are unknown sources.}
\label{figure:tsne_visualize_pointnet}
\end{figure*}

\begin{figure*}[!ht]
\centering
\begin{subfigure}{0.3\textwidth}
\centering
\includegraphics[width=\textwidth]{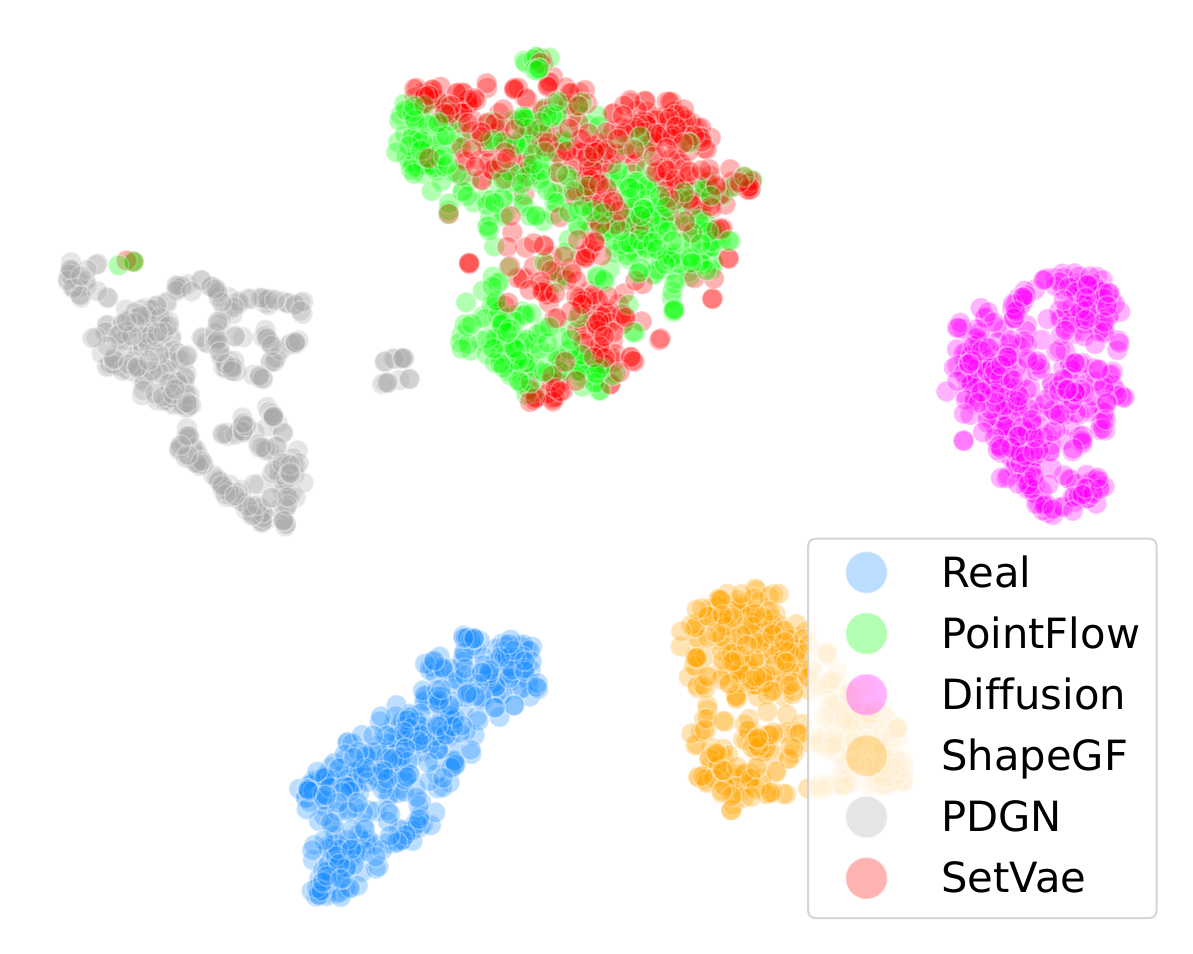}
\caption{Airplane}
\end{subfigure}
\hfill
\begin{subfigure}{0.3\textwidth}
\centering
\includegraphics[width=\textwidth]{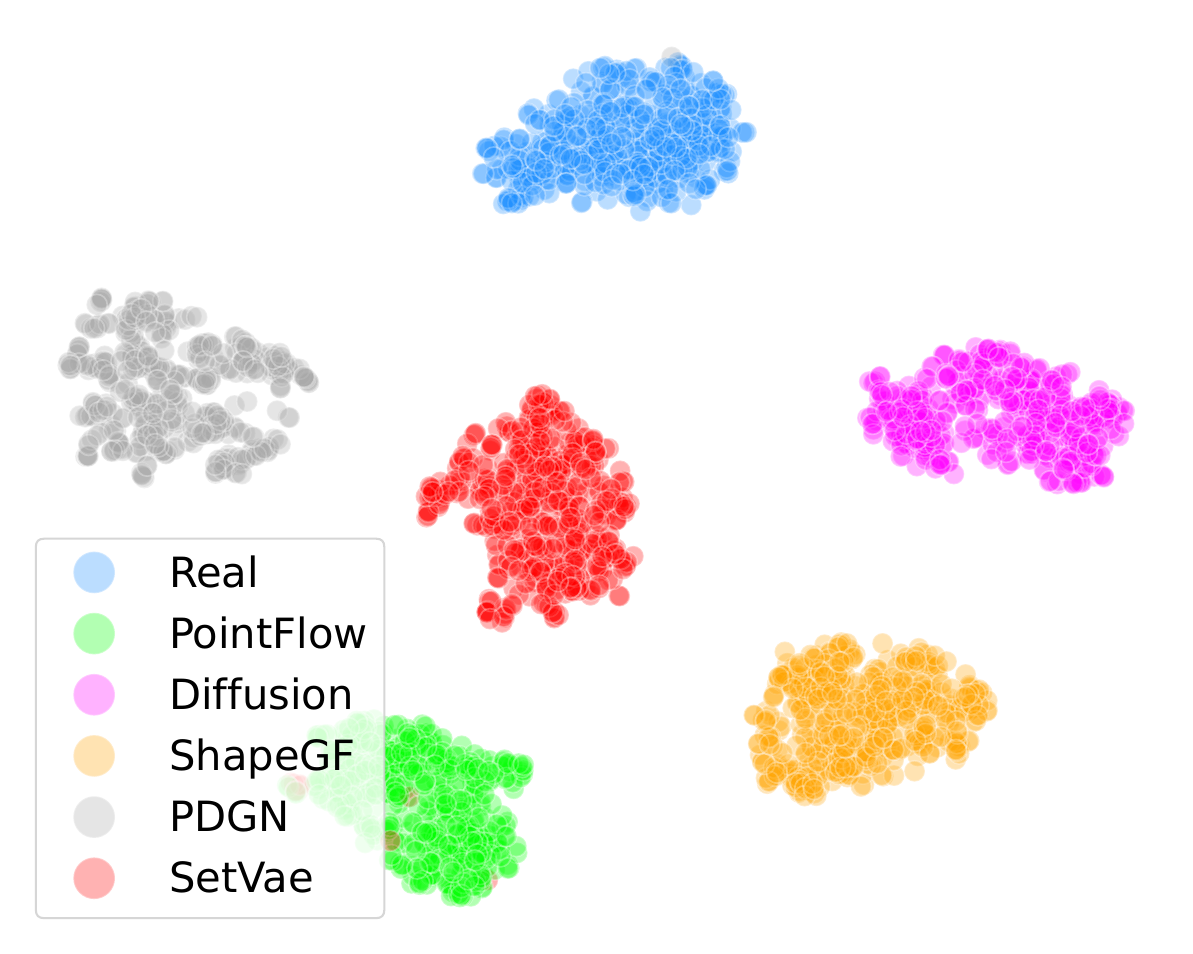}
\caption{Car}
\end{subfigure}
\hfill
\begin{subfigure}{0.3\textwidth}
\centering
\includegraphics[width=\textwidth]{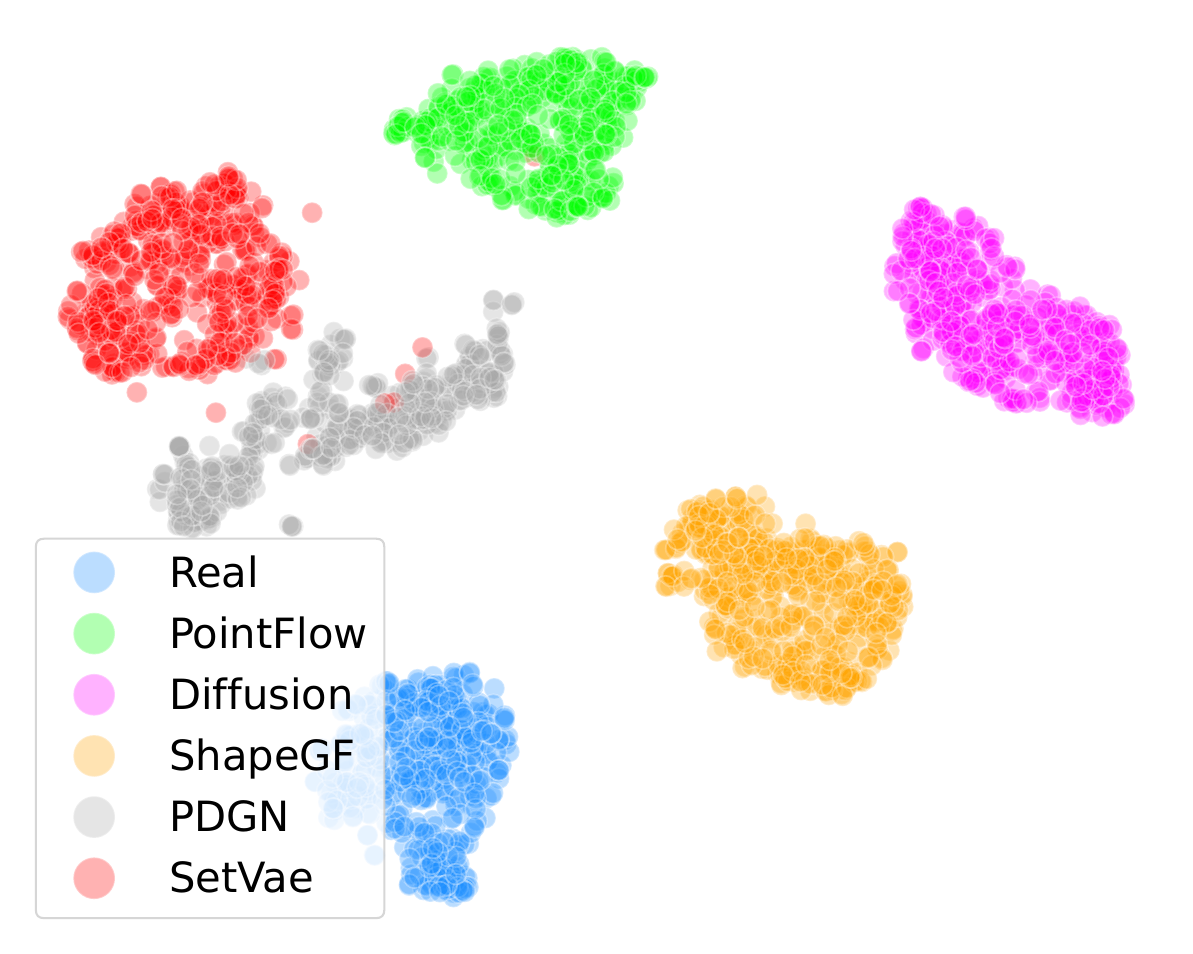}
\caption{Chair}
\end{subfigure}
\caption{t-SNE visualization of clustered point cloud features obtained from \method-DGCNN (better viewed in color).
PDGN and SetVae are unknown sources.}
\label{figure:tsne_visualize_dgcnn}
\end{figure*}

\section{Critical Points Visualization}
\label{appendix: critical_points_visualization}

\begin{figure*}[!ht]
\centering
\begin{subfigure}{0.5\textwidth}
\centering
\includegraphics[width=\textwidth]{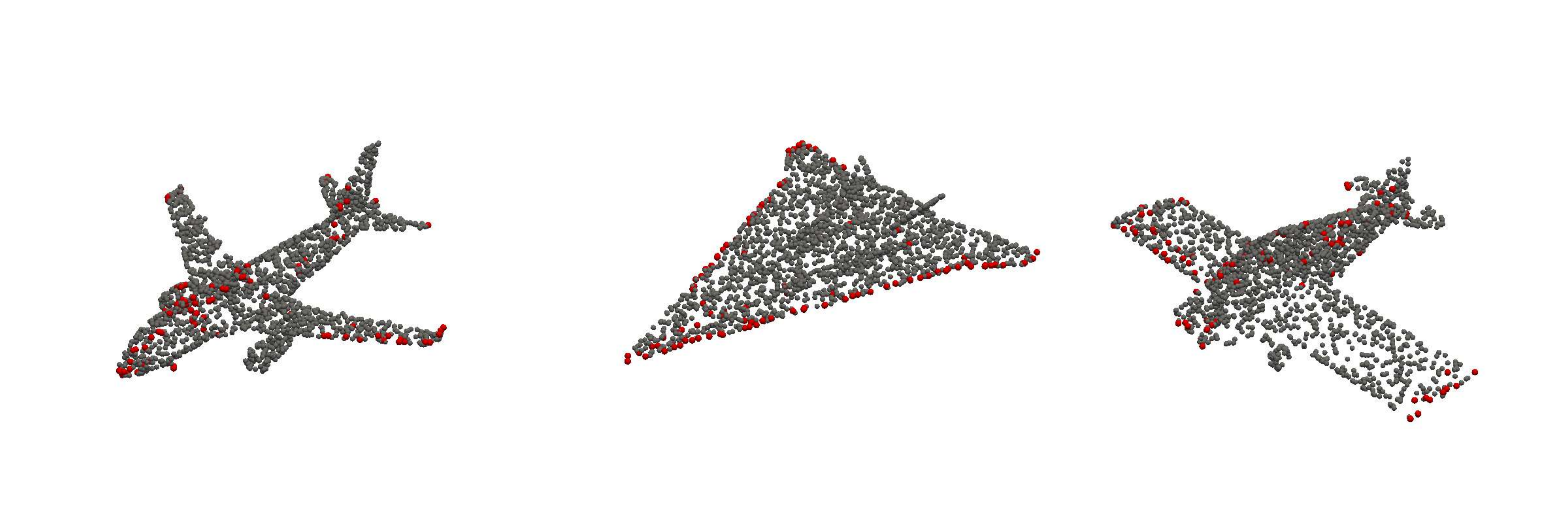}
\caption{Real}
\end{subfigure}
\hfill
\begin{subfigure}{0.5\textwidth}
\centering
\includegraphics[width=\textwidth]{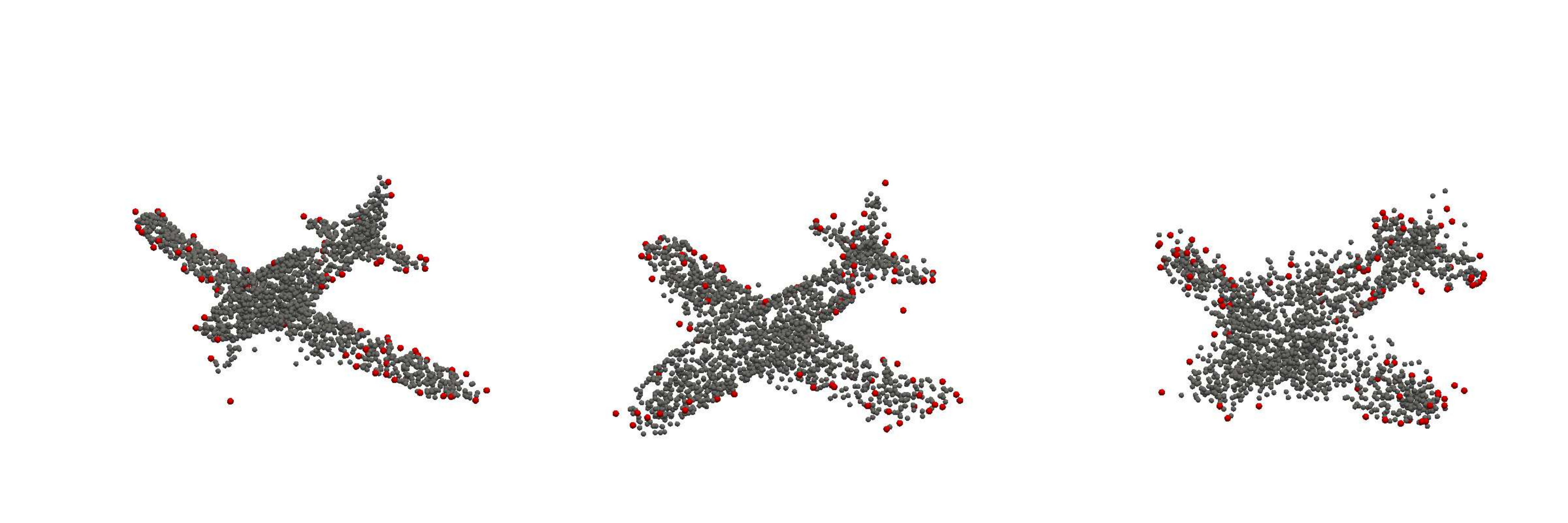}
\caption{PointFlow}
\end{subfigure}
\hfill
\begin{subfigure}{0.5\textwidth}
\centering
\includegraphics[width=\textwidth]{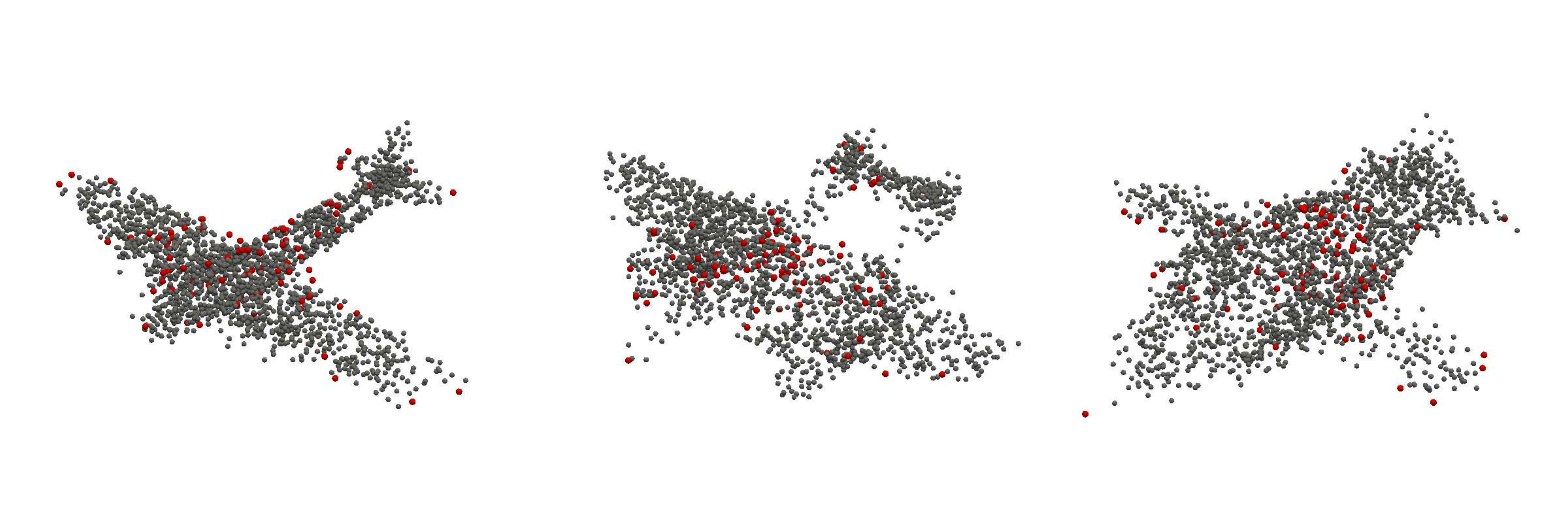}
\caption{Diffusion}
\end{subfigure}
\hfill
\begin{subfigure}{0.5\textwidth}
\centering
\includegraphics[width=\textwidth]{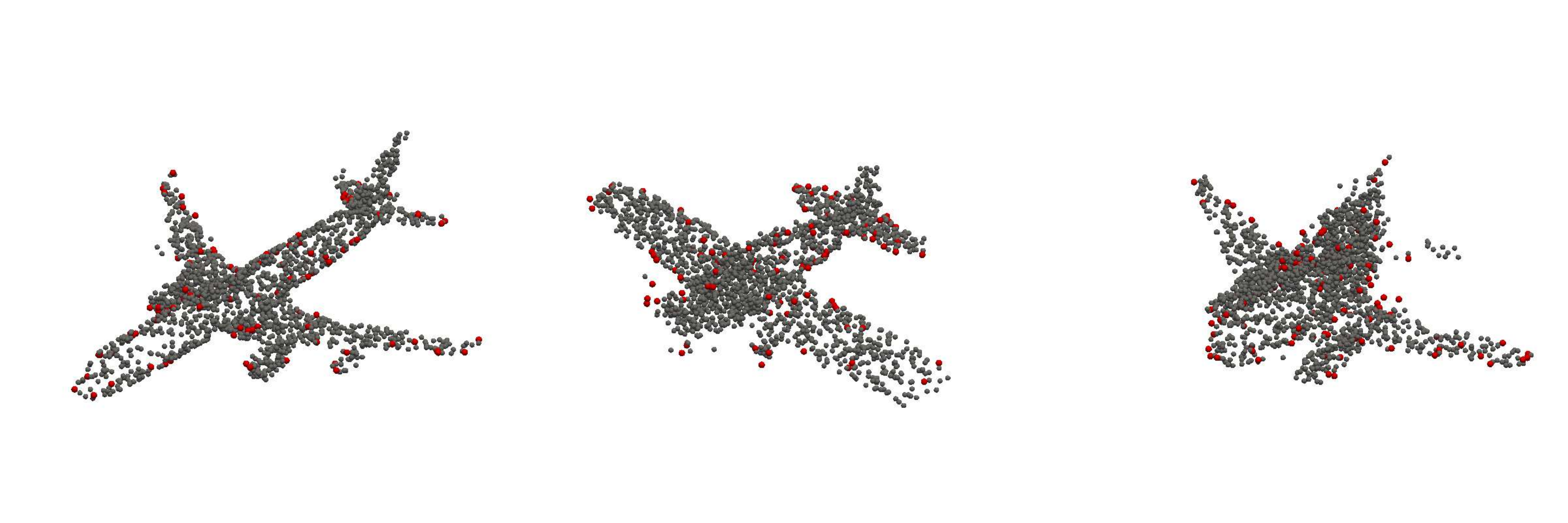}
\caption{ShapeGF}
\end{subfigure}
\hfill
\begin{subfigure}{0.5\textwidth}
\centering
\includegraphics[width=\textwidth]{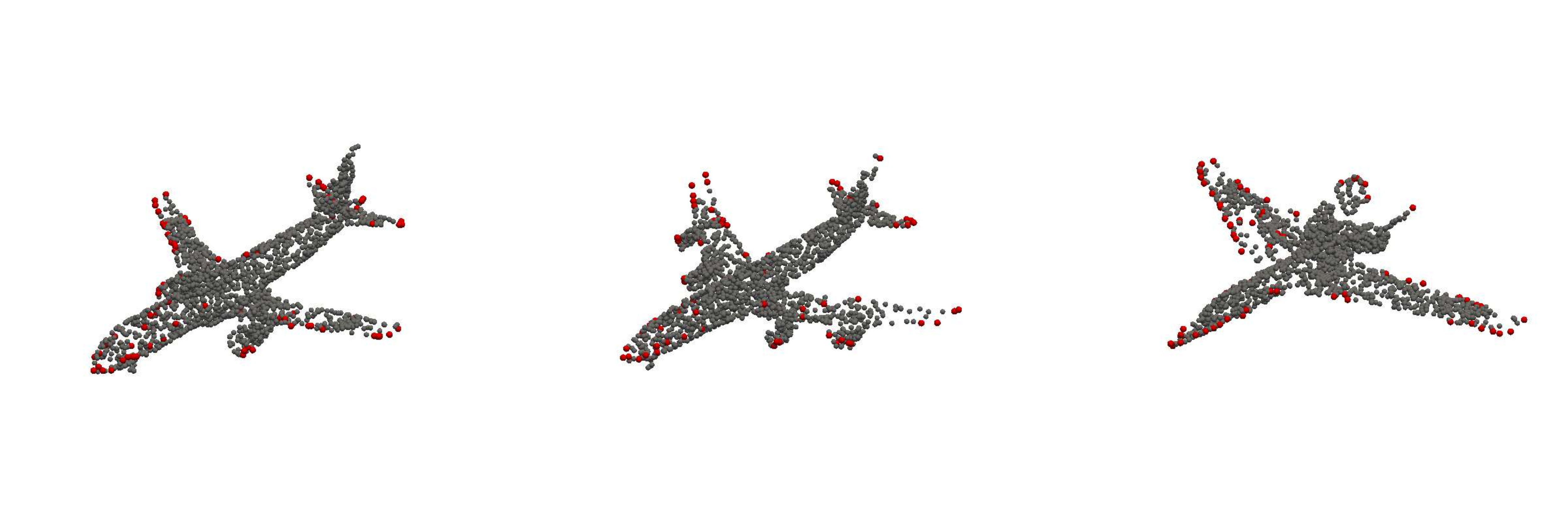}
\caption{PDGN}
\end{subfigure}
\hfill
\begin{subfigure}{0.5\textwidth}
\centering
\includegraphics[width=\textwidth]{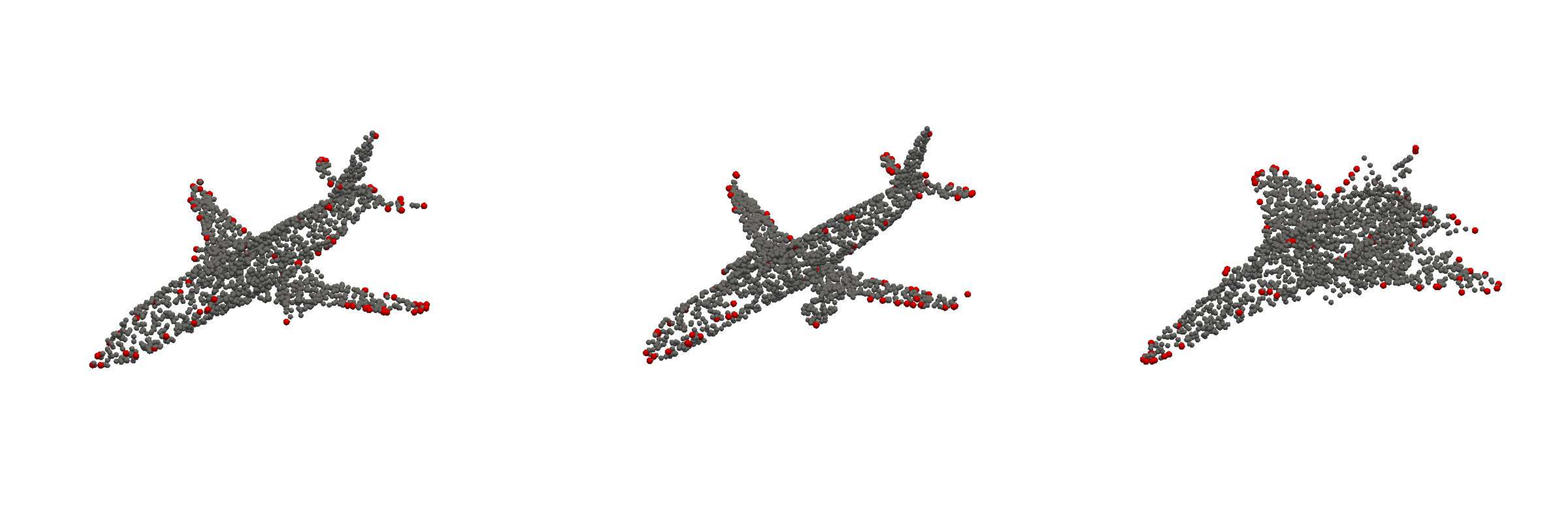}
\caption{SetVae}
\end{subfigure}
\caption{Critical points of Airplane. Grey points describe the original point cloud, and the red points are the identified critical points. 
Each row contains examples from one source and demonstrates certain commonalities.
For example, the majority of critical points focus on the airplane head and wing area in the Real class, while the majority of those critical points are found on the main body in the Diffusion class.}
\label{figure:critical_points_airplane}
\end{figure*}

We explain the intrinsic reason for the source attribution of point clouds in the main text.
\method attributes point clouds from different sources by identifying critical points.
We also extract fingerprints of each source based on the critical points.
\autoref{figure:critical_points_airplane} illustrates the critical points visualization of the shape Airplane when we extract features with PointNet in the \openworld.
For every shape, we select three representative point clouds from each source and also keep point clouds from one column as similar as possible to better compare the critical points distribution.
We use \emph{Chamfer Distance} to find out the similar point clouds in one column.
Point clouds described only by red points are captured by the models, and the information provided by gray points is discarded in the max-pooling operation.
Examples from the same source demonstrate a certain recognition pattern, e.g., in the Real source, critical points primarily describe the head and wings' edge of the airplane; and in the ShapeGF source, these points are scattered at the airplane's surface; in the Diffusion source, extra focus locates in the main body area. 
When we compare the examples from PointFlow and SetVae, their individual patterns are sometimes indistinguishable. 
This observation explains the relatively weaker performance when attributing Airplane.
When we compare airplanes with similar outlooks (smaller Chamfer Distance) in one column, critical points draw slightly different outlines, which explains why \method can recognize point clouds from different sources.

\end{document}